\documentclass[acmlarge,screen,authorversion]{acmart}

\def\BibTeX{{\rm B\kern-.05em{\sc i\kern-.025em b}\kern-.08emT\kern-.1667em\lower.7ex\hbox{E}\kern-.125emX}}
    
\setcopyright{usgovmixed}
\acmJournal{IMWUT}
\acmYear{2019}
\acmVolume{3}
\acmNumber{1}
\acmArticle{1}
\acmMonth{3}
\acmDOI{0000001.0000001}

\usepackage{bm}
\usepackage{booktabs} 
\usepackage{subcaption}
\usepackage[inline]{enumitem}

\usepackage[ruled]{algorithm2e} 

\newcommand{\ra}[1]{\renewcommand{\arraystretch}{#1}}
\newcommand{\system}{\textit{Greina}}

\received{November 2017}
\received[Revised]{May 2018}
\received[Revised]{November 2018}
\received[Accepted]{January 2019}

\begin{document}

\title{Beyond Control: Enabling Smart Thermostats For Leakage Detection} 

\author{Milan Jain}
\orcid{0000-0002-1676-1117}
\affiliation{%
	\institution{IIIT-Delhi}
  	\state{Delhi}
	\country{India}
}

\author{Mridula Gupta}
\affiliation{%
	\institution{University of California Davis}
  	\city{Davis}
	\state{California}
	\country{USA}
}

\author{Amarjeet Singh}
\affiliation{%
	\institution{IIIT-Delhi}
	\state{Delhi}
	\country{India}
}

\author{Vikas Chandan}
\affiliation{%
	\institution{Pacific Northwest National Laboratory}
	\city{Richland}
	\state{Washington}
	\country{USA}
}

\renewcommand{\shortauthors}{Jain, et al.}

\begin{abstract}
Smart thermostats, with multiple sensory abilities, are becoming pervasive and ubiquitous, in both residential and commercial buildings. By analyzing occupants' behavior, adjusting set temperature automatically, and adapting to temporal and spatial changes in the atmosphere, smart thermostats can maximize both - energy savings and user comfort. In this paper, we study smart thermostats for refrigerant leakage detection. Retail outlets, such as milk-booths and quick service restaurants set up cold-rooms to store perishable items. In each room, a refrigeration unit (akin to air-conditioners) is used to maintain a suitable temperature for the stored products. Often, refrigerant leaks through the coils (or valves) of the refrigeration unit which slowly diminishes the cooling capacity of the refrigeration unit while allowing it to be functional. Such leaks waste significant energy, risk occupants' health, and impact the quality of stored perishable products. While store managers usually fail to sense the early symptoms of such leaks, current techniques to report refrigerant leakage are often not scalable. We propose \system~ - to continuously monitor the readily available ambient information from the thermostat and timely report such leaks. We evaluate our approach on $74$ outlets of a retail enterprise and results indicate that \system~ can report the leakage a week in advance when compared to manual reporting.
\end{abstract}

%
%


%
%



\thanks{This work was done in collaboration with Zenatix Solutions Pvt. Ltd. We would like to thank TCS Innovation Lab for Ph.D. fellowship to the first author, the administration of the retail enterprise to allow us to collect data, and all the outlet managers and the technicians to help us during the data collection.}

\maketitle

\renewcommand{\shortauthors}{M. Jain et al.}

\section{Introduction}
\label{sec:introduction}
\begin{figure}[tp]
    \begin{minipage}[t]{0.32\linewidth}
        \centering
        \includegraphics[width=\linewidth]{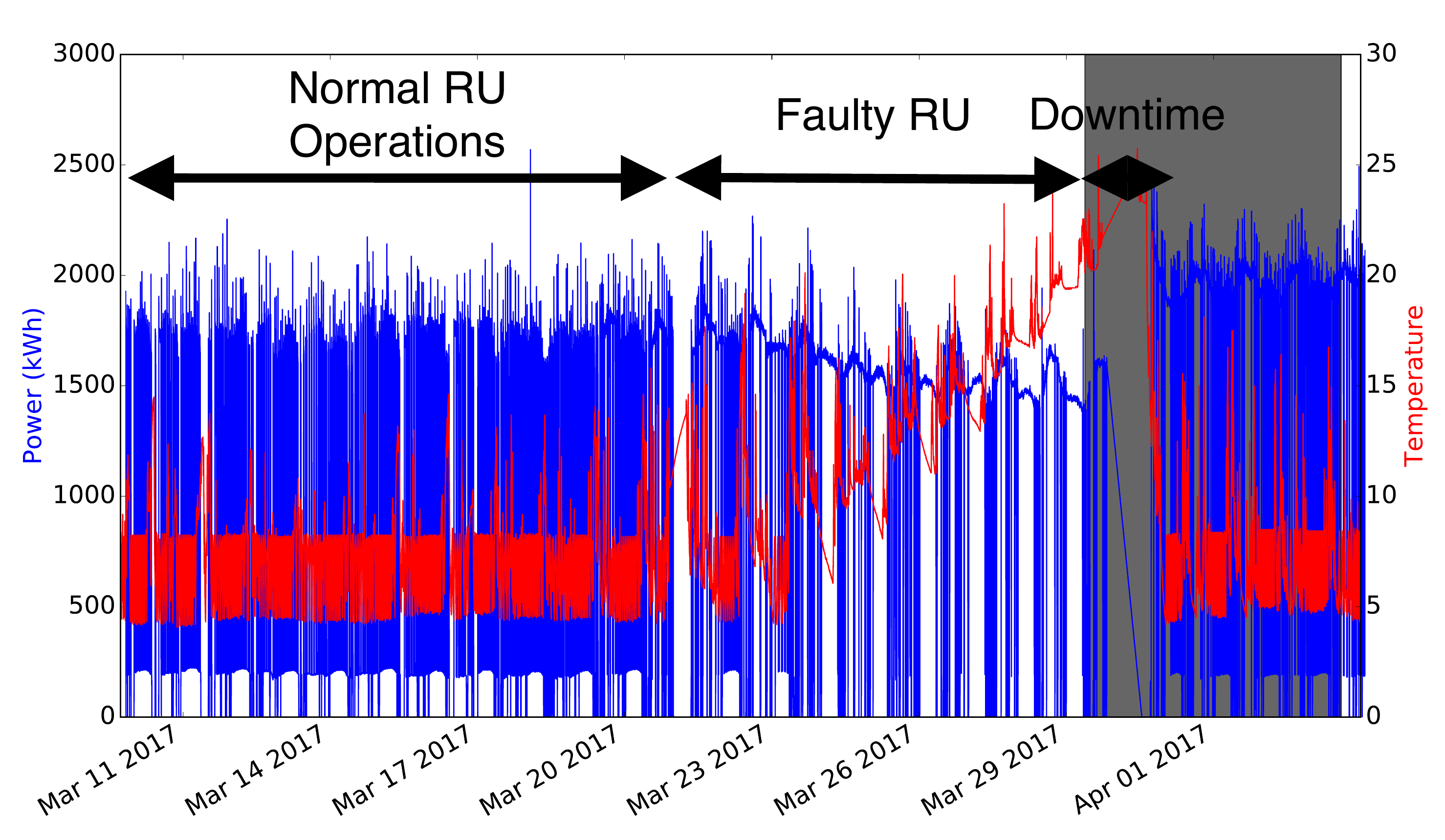}
        \subcaption{Gas leakage started on March 21, but manager kept using the RU for a week and complained on March 29 when it stopped cooling.}
        \label{fig:faulty_ru}
    \end{minipage}
    \hspace{0.01\linewidth}
    \begin{minipage}[t]{0.32\linewidth}
        \centering
        \includegraphics[width=\linewidth]{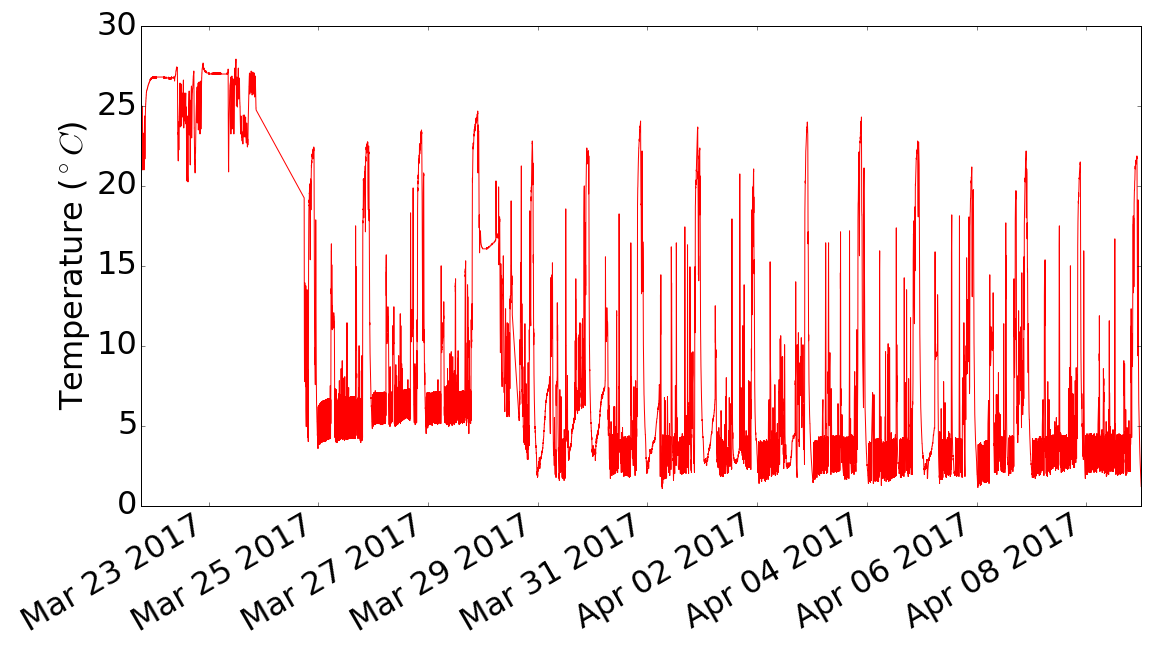}
        \subcaption{During daytime store manager frequently visits the cold-room which leads to high temperature in working hours.}
        \label{fig:noise_occ}
    \end{minipage}
    \hspace{0.01\linewidth}
    \begin{minipage}[t]{0.32\linewidth}
        \centering
        \includegraphics[width=\linewidth]{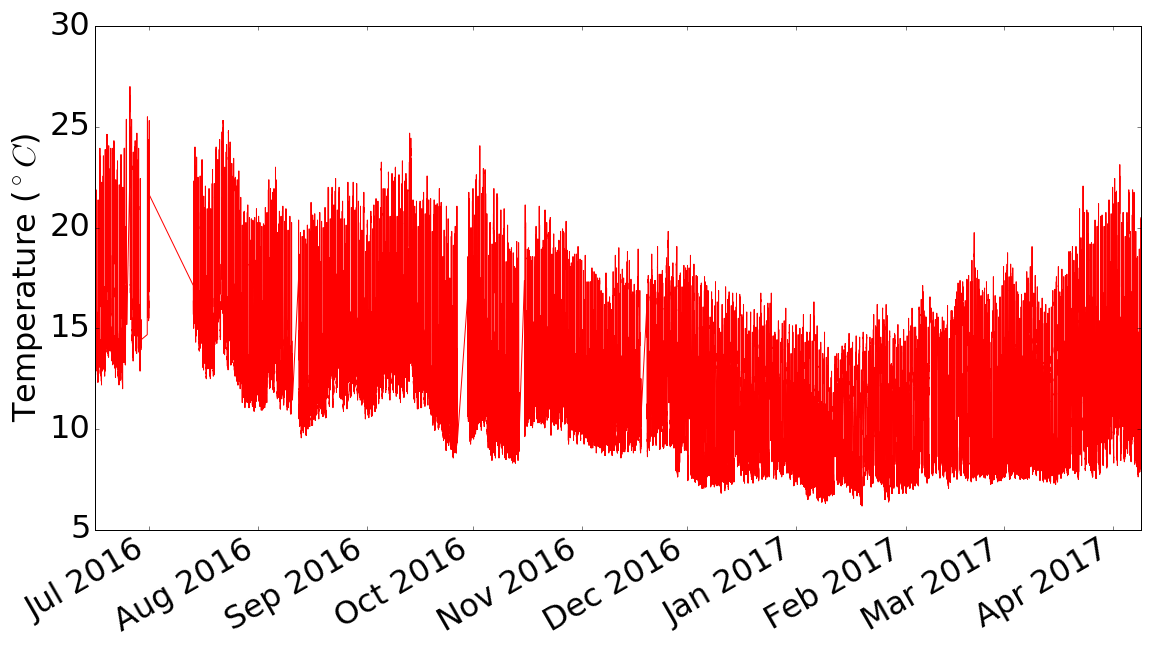}
        \subcaption{In winters (December 2016 - March 2017), room temperature stays in a lower range as compared to summers.}
        \label{fig:weather_1}
    \end{minipage}
    \caption{[Left] Delay in reporting refrigeration leak led to complete shutdown of store operations for 2-3 days. [Middle and Right] Room temperature depends on the activities of store manager and seasonal environmental changes.}
\end{figure}

Refrigerant leakage is a common mechanical fault in compressor based appliances which are primarily used for air-conditioning and refrigeration purposes~\cite{brambley2005advanced}. The puncture hole often starts as a pinhole leak and becomes bigger when goes undetected. Due to the loss of refrigerant, compressor (a component within the RU) works with reduced efficiency and takes more time than the usual to cool the room; thus, wasting significant energy. In addition to that, the leakage exposes tenants to refrigerant which is extremely dangerous for their health~\cite{assawamartbunlue2006refrigerant}. The consequences of refrigerant leakage are even worse for retail outlets who set up cold-rooms to preserve perishable food items (usually at $5^\circ$C-$8^\circ$C), and the stored product goes bad due to improper cooling by the RU during the leakage. Early detection of such leaks can benefit the retail enterprises in - 
\begin{enumerate*}
    \item increasing their profits by reducing the energy wastage,
    \item avoiding leakage of hazardous refrigerant in the open environment, and
    \item maintaining the product quality~\cite{downey2002can, dong2013integrated, francis2017investigation, davis2015contribution}.
\end{enumerate*}

Unfortunately, store managers usually lack essential skills to sense the leakage at an early stage, and undetected leaks widen with time as RU remains functional during the leakage. At breakdown point, RU stops cooling followed by a complete shutdown of the outlet for several days. Figure~\ref{fig:faulty_ru} depicts one such scenario where store manager reported the problem when RU broke, and the store went out of operations. In addition to repairing cost, the owner also dealt with the staling of stored products and business loss owing to the downtime. Current techniques for leakage detection are either direct or indirect~\cite{yoo2017refrigerant}. Direct methods, such as Halide leak detector, require an expert to use the device with certain precautions for on-spot leakage detection~\cite{bender2003selective, jeffers1984halogen, martell1994refrigerant}. On the other hand, indirect techniques need specialized sensors to monitor temperature, pressure, and mass flow rate at multiple points inside the appliance~\cite{fisera2012system, ren2008fault, payne2015self}. As pressure and mass flow sensors are more expensive than temperature sensor, several studies focused on limiting the number of sensors for leakage detection~\cite{fisera2012performance, yoo2017refrigerant}. However, even with the limited sensors, Yoo et al.~\cite{yoo2017refrigerant} required temperature sensors at five points inside the system - 
\begin{enumerate*}
    \item air inlet of indoor unit, 
    \item air inlet of outdoor unit, 
    \item evaporator midpoint, 
    \item condenser midpoint, and
    \item compressor discharge point.
\end{enumerate*}
Such an extensive sensor installation (that too within the appliance) often limits indirect techniques to laboratory setup and considered as an expensive and unscalable solution to the problem.

In this paper, we reinforce smart thermostats with \system\footnote{\system~ is an \emph{Icelandic} word that means \emph{to identify.}} - a framework that observes deviations in the measured temperature from the estimated temperature to detect refrigerant leakage. The estimates come from a lumped parameter thermal model whose parameters are tuned to a particular room environment by the ambient information sensed by the thermostat. Smart thermostats are 
\begin{enumerate*}
    \item plug-n-play - no need of technician for installation, 
    \item allow remote sensing of RU, 
    \item never intervene in the daily routine of the store managers, and
    \item anticipated to increase by 400\% in next couple of years~\cite{leuth2018smart}.
\end{enumerate*}
Given the benefits, studies in the past explored smart thermostats for user feedback~\cite{jain2016non}, occupancy detection~\cite{kim2017improved}, energy-efficient control~\cite{lu2010smart}, among many other applications. However, to the best of our knowledge, no one has studied smart thermostats for leakage detection. We believe enhancing smart thermostats for leakage detection presents a genuinely low-cost and scalable solution for the problem. 

Despite all the benefits, one must also note that the overall problem is non-trivial and exhibits multiple challenges. Room temperature is sensitive to the actions of outlet manager (Figure~\ref{fig:noise_occ}) and outside weather conditions (Figure~\ref{fig:weather_1}), and both these parameters differ significantly across the outlets. Even within a store, the daily routine of the manager, climatic conditions, and fitness level of an appliance change considerably with time. To consider both temporal and spatial variations, \system~ utilizes a first-order thermal model which simulates the room temperature while considering the influence of outlet manager and climate outside the room. \system~ collects weather data through a cloud-based weather service, and monitors user activity through a door sensor. With time, as the framework receives more and more \emph{clean} data, it keeps updating the model parameters to accommodate temporal changes in the building's thermal behavior. To gather \emph{clean} data for the learning phase, \system~ does not consider those days which were identified as leaking while monitoring the RU. In the initial stages, when sufficient \emph{clean} data is unavailable to tune the model parameters, \system~ benefits from transfer learning and uses the parameters of a contextually `similar' outlet to simulate the room temperature. Transfer learning ensures that the proposed framework is ready-to-work from the day of installation. Eventually, \system~ compares the estimates from the tuned thermal model with sensor measurements and when the actual temperature is sufficiently above the estimate, the system raises a red flag.

Of course, not every red flag is a leakage flag. Room temperature can be higher than the estimates even when RU is working fine (false positive), and vice-a-versa, the framework might confuse the initial symptoms of leakage with the noise generated from manual interventions (false negative). For instance, in Figure~\ref{fig:faulty_ru}, the room temperature is in the range of $8^\circ$C-$10^\circ$C in the initial stages of the refrigerant leak, akin to temperature variations due to tenants' activities in Figure~\ref{fig:noise_occ}. If misclassification is a false negative (misinformed that RU is fine), then repairing will get delayed until the outlet manager identifies the leakage. However, if misclassification is a false positive (misinformed that RU broke), then the company might end up paying a significant amount to their maintenance contractor for the unnecessary visits. While such visits are immoderate, they are annoying for the store manager and disruptive to the daily operations. Henceforth, \system~ employs CUSUM (Cumulative Sum Control) technique to ensure that the user gets notified only when the system is confident about the leakage.

We evaluate \system~ using the data collected from 74 outlets of a retail enterprise for one year. For ground truth verification, we use fault logs as reported by the servicing company after addressing the fault. Our results indicate that \system~ is comparable or better than manual reporting. The proposed framework can reduce average reporting delay by 5-6 days, when compared to manual reporting. In other words, even if outlet managers could use the leakage detectors (such as Halide leak detector), they would have delayed the repairing by almost a week. We couldn't compare \system~ with indirect methods (of leakage detection) because proposed techniques require specialized sensors to monitor temperature at multiple points within the RU.

Given the energy-saving features and other advancements, smart thermostats are anyways going to replace traditional thermostats very soon. As of now, thermostats can monitor room temperature, tenants' daily activities~\footnote{Though, in this study, we used door sensor to monitor human activities, Appendix~\ref{app:extension} discusses an extended thermal model which replaces door sensor with the motion sensor of the smart thermostat.}, and even upload the sensed information to the cloud. Location-based detailed weather information is readily available from several cloud-based climate monitoring applications. By combining the two pieces of information, we can use \system~ for leakage detection without instrumenting the appliance. In addition to being scalable, the proposed approach is reliable enough to maintain food quality and generate minimal false alarms. Finally, \system~ requires minimal (or no) intervention from the store manager and ensures non-interruptive working hours. Currently, in collaboration with an energy-analytics based venture, we are working on the deployment of proposed thermostats across all the 74 outlets which were considered as part of this study.

\section{Related Work}
\label{sec:relatedwork}
Anomaly detection is an active area of research with extensive literature in numerous domains and applications~\cite{chandola2009anomaly, kim2018review}. One such problem is fault detection and diagnosis (FDD) where the objective is first to identify a malfunctioned appliance followed by a \emph{root-cause} analysis to diagnose the problem~\cite{gertler2013fault, isermann2011fault}. Usually, the partial or complete failure of the hardware (present in the equipment) brings down the whole system. Though the literature on fault detection in mechanical units is considerably extensive, we will limit the scope of this survey only to compressor-based appliances.

\subsection{Fault Detection Frameworks}
Typically, fault detection frameworks for compressor-based appliances are designed either for Refrigeration Units~\cite{behfar2017automated, thybo2004development} or Heating, Ventilation, and Air Conditioning Units (HVAC)~\cite{wirz2017commercial, brambley2005advanced, diagnostics2005advanced, yu2014review}. Irrespective of the type of appliance, most fault detection engines are based on a conventional design - simulate a baseline and compare with the measured signal (mostly energy) to mark any deviations due to the fault~\cite{du2007fault, meng2006rare, narayanaswamy2014data, ganu2014socketwatch, palani2014putting, chen2001simple}. O'Neil et al.~\cite{o2014model} simulated the energy consumption through a building simulation framework, EnergyPlus~\cite{crawley2000energy}, and compared with actual energy consumption data to identify unacceptable performance. Mavromatidis et al.~\cite{mavromatidis2013diagnostic} computed baseline energy consumption through an artificial neural network model and used the baseline to detect faults in supermarket refrigerators. Li et al.~\cite{li2014experimental} studied the correlation between electrical signals and the common faults for roof-top air conditioners by conducting a series of experiments. Srinivasan et al.~\cite{srinivasan2015bugs} proposed an energy model to pick anomalies in the power signature of supermarket refrigerators.

Studies also monitored different other parameters to develop a black box model for fault detection~\cite{yang2011afault, yang2011bfault, han2010pca}. Keres et al.~\cite{keres2016fault} observed compressor frequency to identify a faulty refrigerant unit. Porter et al.~\cite{porter2008refrigeration} collected actual operating parameters from a set of microsystem sensors installed throughout the refrigeration and compared with ideal conditions of the system for fault detection. Payne et al.~\cite{payne2015self, payne2018data} monitored temperature at nine points within an air-conditioner for a self-trained fault-free model and then a data-clustering approach to segregate faulty instances. Similarly, Kulkarni et al.~\cite{kulkarni2018predictive} designed a random forest binary classifier to detect issues in refrigeration cases by observing temperature and defrost state within supermarket refrigeration cases. While most of the above-mentioned studies focused only on detecting `abrupt' faults, none of them looked into the problem of refrigerant leakage. Though such a fault detection engine would confirm the presence of a fault, the engine won't label the fault. In the absence of labeling, we couldn't compare Greina with any of the above-mentioned generic fault detection frameworks\footnote{If we would have considered all anomaly instances as leakage, our analysis would have shown high false alarm rate for any generic fault detection framework. Therefore, though we had data to implement some of these techniques, we couldn't compare the performance of any generic fault detection framework with Greina in our evaluation (Section~\ref{sec:evaluation}).}.

\subsection{Leakage Detection Frameworks}
Specifically for leakage detection, exiting systems are typically categorized as direct and indirect methods.

\subsubsection{Direct Methods}
In direct methods, typically a technician visits the site and uses a leakage detector~\cite{bender2003selective, martell1994refrigerant} to confirm the leak. Hailey detector~\cite{jeffers1984halogen} is one such widely used leak detector in which flame changes the color when the refrigerant is present in the environment. Parekh~\cite{parekh1992method} proposed composition of fluorescent, alkyl substituted perylene-dye, and a polyhalogenated hydrocarbon refrigerant to visually detect refrigerant leakage. Though these detectors are accurate, one must use them with certain precautions. For instance, we cannot use Hailey detector with hydrocarbon refrigerants. Moreover, the effectiveness of direct methods primarily relies on manager's ability in timely detecting the leakage.

\subsubsection{Indirect Methods}
Alternatively, in indirect methods, studies often monitor multiple parameters at different points within the appliance for leakage detection~\cite{jeong2005refrigerant, tassou2005fault, rinehart2004refrigerant, kadle2014refrigerant, dube2009refrigerant, morrow1994refrigerant, morrow1996refrigerant}. Taylor et al.~\cite{taylor2004refrigerant} designed a neural network to analyze data from multiple monitoring alarms to predict refrigerant leakage. Rossi et al.~\cite{rossi1997statistical} proposed a statistical rule-based leakage detection technique which could detect 5\% loss of refrigerant through extensive instrumentation. Breuker et al. \cite{breuker1998common} studied temperature at nine different locations and relative humidity to characterize \emph{soft} faults (such as loss of refrigerant) and their impact on the operations of rooftop air conditioners. Since the majority of leakage detection studies~\cite{li2004decoupling, navarro2006vapour, grace2005sensitivity} assume steady-state, Kim et al.~\cite{kim2008design} proposed a methodology for developing a steady-state detector for any generic vapor-compression system. Steady-state in itself is a misleading term because parameters of a vapor-compression system are dynamic in real-world~\cite{kim2008design}. Therefore, the study recommended inclusion of all leakage detection features (for best results) which makes the overall approach expensive due to extensive instrumentation of the appliance.

To minimize the number of sensors and sensing points, Fisera et al.~\cite{fisera2012system, fisera2012performance} developed an energy consumption model by monitoring nine parameters (including relative humidity and carbon-dioxide level) to distinguish anomalous and degradation events. In a patented technology, Suzuki et al.~\cite{suzuki2004vapor} compared theoretical heat dissipation of the condenser with the actual temperature difference (in the condenser) for leakage detection. In the same direction, Yoo et al.~\cite{yoo2017refrigerant} monitored temperature difference between inlet air and midpoint of the heat exchanger to detect the refrigerant charge level. Though stated approaches require less number of sensing points, invasive sensing often limits their evaluation to a controlled environment. Consequently, the efficacy relies on the validity of assumptions from the controlled environment to a real-world scenario.

\section{Approach}
\label{sec:approach}
\begin{figure*}[t!]
    \begin{minipage}[t]{0.6\linewidth}
        \centering
        \includegraphics[width=\linewidth]{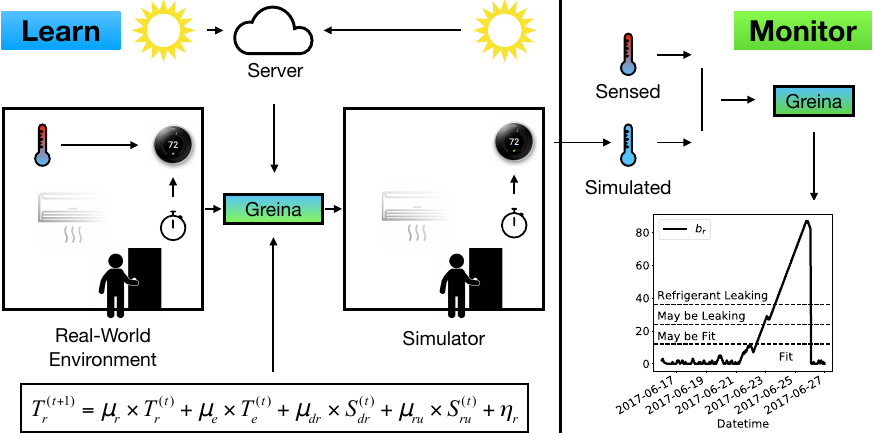}
        \subcaption{\system~ is a two-step process. The proposed framework first learns the typical temperature profile of the room followed by monitoring the RU for refrigerant leakage.}
        \label{fig:system_architecture}
    \end{minipage}
    \hspace{0.01\linewidth}
    \begin{minipage}[t]{0.38\linewidth}
        \centering
        \includegraphics[width=\linewidth]{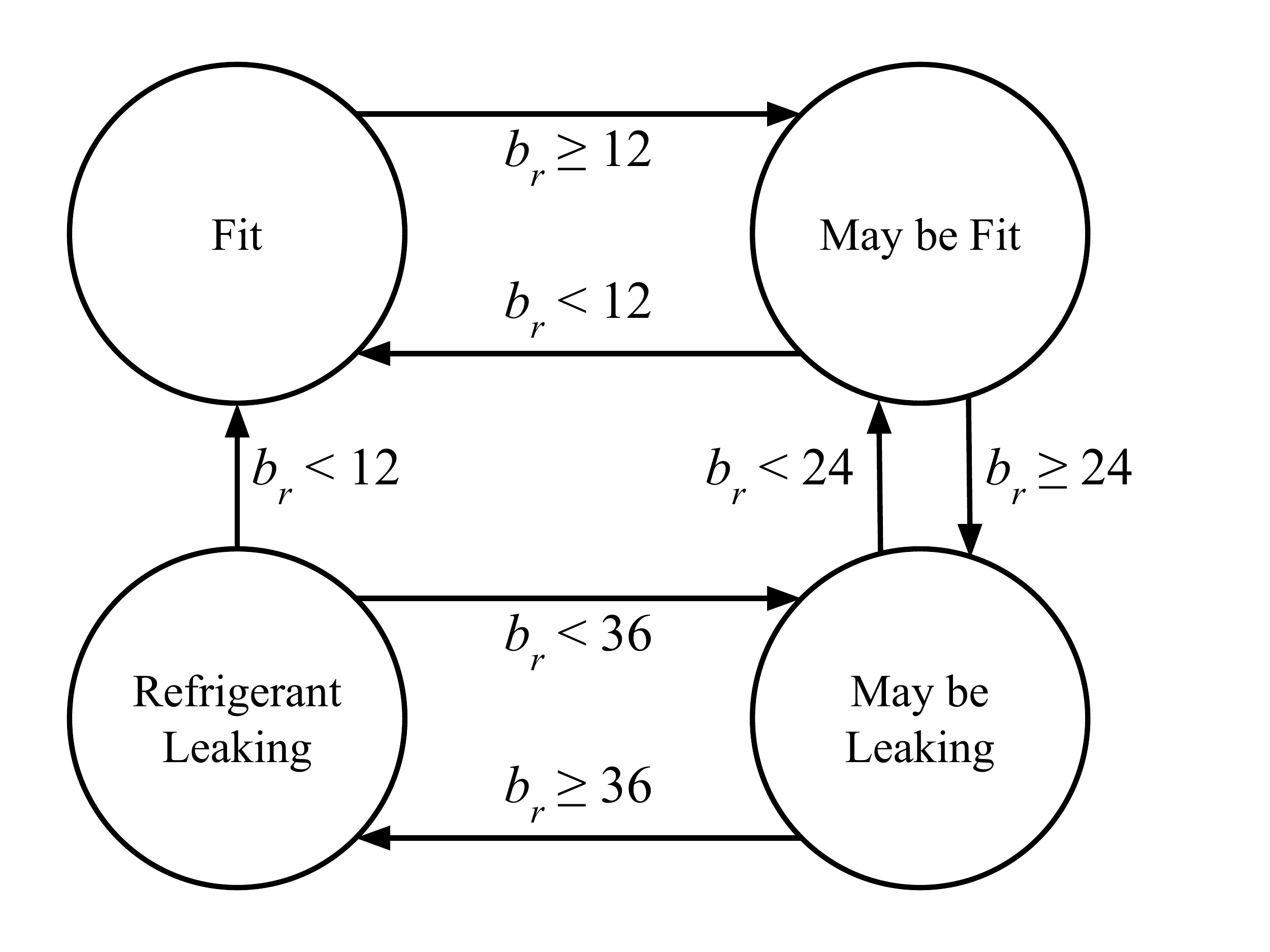}
        \subcaption{For each room $r$, \system~ maintains a bucket variable ($b_r$) to label the RU at each hour.}
        \label{fig:state_identification}
    \end{minipage}
    \caption{[Left] \system~ takes ambient information from the smart thermostat and weather conditions from a third-party cloud-based weather server to tune the parameters of a lumped thermal model. Tuned thermal model simulates the room temperature for leakage detection. To incorporate temporal changes in the environment, \system~ regularly updates the model parameters. [Right] From data, we observed that if the temperature within a room is consistently beyond the estimates for 36 hours, then there are high chances of refrigerant leakage in the RU.}
    \label{fig:greina_system}
\end{figure*}

In \system, every outlet goes through a two-step process - 
\begin{enumerate*}
	\item learn the \textit{normal} behavior, and
	\item monitor refrigeration unit for leakage, 
\end{enumerate*}
as shown in Figure~\ref{fig:greina_system}.

\subsection{Learn Normal Behaviour}
In a typical room $r$, the change in temperature, in a given time interval $\tau$ (in seconds), primarily depends on the heat transferred by the weather outside the room, the heat added by the open door, and the heat extracted by the refrigeration unit (Equation~\ref{eq:thermal_model}).

\begin{equation}
    \frac{(T_r^{(t+1)} - T_r^{(t)}) \times C_r}{\tau} = K_e^r \times (T_e^{(t)} - T_r^{(t)}) + (Q_{dr} \times S_{d_r}^{(t)}) + (Q_{ru} \times S_{ru_r}^{(t)}) + \eta_r
    \label{eq:thermal_model}
\end{equation}

Here, $T_r^{(t)}$ and $T_e^{(t)}$ denote the average room and external temperature in the last time interval $\tau$ (between $t$ and $t-1$), respectively. $C_r$ is the thermal capacity of the room, and $K^r_e$ is the heat transfer coefficient between room and external weather conditions. We derived the model from Bacher et al. \cite{bacher2011identifying}.

When supply in the front runs out, the manager opens the room to get the fresh stock and $S^{(t)}_{d_r}$ denotes the state of door (open/close) at time $t$. Given that RU is a two-state appliance - 
\begin{enumerate*}
    \item compressor on, and 
    \item compressor off, 
\end{enumerate*}
where compressor is the major power consuming component of RU, $S^{(t)}_{ru_r}$ is the state of RU at time $t$. Correspondingly, $Q_{dr}$ and $Q_{ru}$ denote the amount of heat added through the door, and heat extracted by the refrigeration unit, respectively. $\eta_r$ is the thermal noise introduced by the random events.

\begin{equation}
    \begin{split}
    \mu_r = 1 - \frac{K_e^r * \tau}{C_r}, \quad &
    \mu_e = \frac{K_e^r * \tau}{C_r}, \quad
    \mu_{dr} = \frac{Q_{dr} * \tau}{C_r}, \quad \\
    \mu_{ru} = \frac{Q_{ru} * \tau}{C_r}, \quad &
    \eta_r^\prime = \frac{\eta_r * \tau}{C_r}
    \end{split}
    \label{eq:lumped_parameters}
\end{equation}

In state space representation, we can lump the parameters (as shown in Equation~\ref{eq:lumped_parameters}) and rewrite the thermal model as Equation~\ref{eq:lumped_model}. \system~ learns the model parameters ($\theta = <\mu_r, \mu_e, \mu_{dr}, \mu_{ru}, \eta^\prime_r>$) through linear regression while using the data streams of room temperature ($\mathbf{T_r}$), external temperature ($\mathbf{T_e}$), door status ($\mathbf{S_{d_r}}$), and RU status ($\mathbf{S_{ru_r}}$). Tuning the parameters of a physics-based thermal model by using the real-world data is called \emph{Grey Box Modelling} and widely practised by the researchers in several domains~\cite{dewson1993least}. As tuning involves sensor data, the model with adjusted parameters represents an approximate thermal behaviour of the room.

\begin{equation}
    T_r^{(t+1)} = \mu_r \times T_r^{(t)} + \mu_e \times T_e^{(t)} + \mu_{dr} \times S_{d_r}^{(t)} + \mu_{ru} \times S_{ru_r}^{(t)} + \eta^\prime_r
    \label{eq:lumped_model}
\end{equation}

\system~ tunes model parameters every month through Stochastic Gradient Descent~\cite{pedregosa2011scikit}. The online learning ensures that model adapts to recent changes occurring in the environment, without forgetting the existing knowledge. Now, to tune the model parameters, the proposed framework require \emph{clean} data - data when the refrigeration unit was working fine. To gather \emph{clean} data for the learning phase, \system~ does not consider those days which were identified as leaking during the monitoring phase. For instance, if \system~ labeled refrigeration unit as \emph{Refrigerant Leaking} for five days in the last month, the learning module won't consider data from those five days to tune the model parameters. The rationale behind this is that if RU was unable to keep room temperature within limits for more than 36 hours consecutively, it was not an ideal situation and there was a problem. During those five days, the problem could have been Refrigerant Leakage (true positive), or an ongoing maintenance (false positive); either way, those days are not ideal (or regular) days for the training. Thus, we discard those from the learning data and only use \emph{clean} data.

Initially, we specified two key features of \system~ - 
\begin{enumerate*}
    \item it uses readily available information from the smart
thermostat, and 
    \item it can start monitoring the RU without waiting for the sensor data for a long time. 
\end{enumerate*}    
Though the proposed framework can collect room temperature ($\mathbf{T_r}$) and door status ($\mathbf{S_{d_r}}$) information from the thermostat, and external temperature ($\mathbf{T_e}$) from any third-party weather server, the refrigeration state ($\mathbf{S_{ru_r}}$), which typically requires appliance-level monitoring is unavailable to the framework. In addition to that, waiting for \emph{clean} data for a long time in the initial stages of deployment can significantly delay the monitoring process. \system~ leverages domain advancements to fulfill these requirements.

\begin{figure*}
    \begin{minipage}{0.49\linewidth}
		\begin{algorithm}[H]
			\SetAlgoNoLine
			\tcc{$D^h_r$ indicates the median of total number of door opening instances occurred in the $h^{th}$ hour for the $r^{th}$ room across all the days. $h \in [1, 24]$}	
			\KwIn{$\mathbf{D_r}$ where $r$ is the ID of new room}
			\KwOut{Parameters ${<\mu_r, \mu_e, \mu_{dr}, \mu_{ru}, \eta^\prime_r>}$ of most `similar' outlet (room $r^*$)}
			
			\BlankLine
			\tcc{Score indicates the `similarity' in the door opening patterns of two rooms across the day.}
			\textbf{Compute Score:}
			
			$S_{i} = \lVert \mathbf{D_{i}} - \mathbf{D_{r}} \rVert^2$, $\forall i \in $~\{available outlets\}\;
			
			\BlankLine
			\tcc{Select the room number with minimum score}
			\textbf{Select `Similar' Outlet:}
			
			$r^* = \operatornamewithlimits{argmin}\limits_{i}(\mathbf{S})$
			\caption{Ranking Algorithm}
			\label{algo:ranking}
		\end{algorithm}
    \end{minipage}
    \hspace{0.01\linewidth}
    \begin{minipage}{0.49\linewidth}
    	\includegraphics[width=\linewidth]{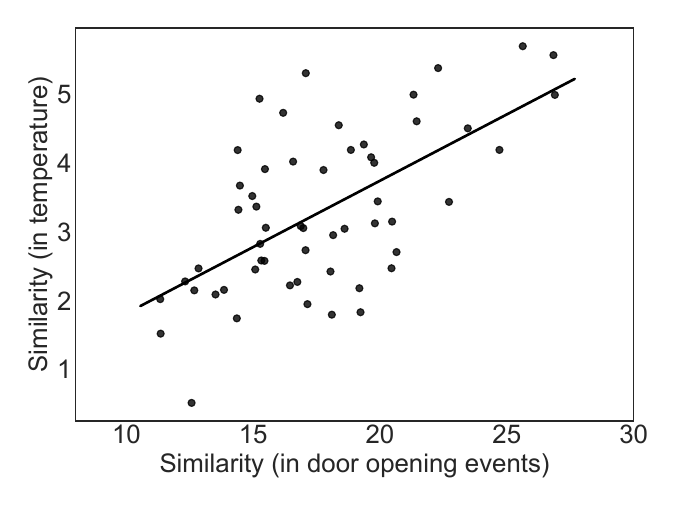}
       	\caption{If the managers of two rooms have `similar' routine, then there are good chances that the hourly temperature profile of those rooms will also be `similar'.}
        \label{fig:similarity_oplot}
    \end{minipage}
\end{figure*}

\subsubsection{State of Refrigeration Unit}
Jain et. al.~\cite{jain2016non}, proposed a classification based algorithm to determine compressor state from the room temperature. The intuition was to first identify compressor on and off events from the temperature data followed by event sequencing to estimate the state vector for RU. When compressor turns on, the room temperature quickly goes down due to the addition of cold air, and when compressor turns off, the room temperature shoots up due to thermal leakage. The estimation algorithm (EA) uses \emph{k-means} to segregate sudden increase and sudden decrease in room temperature (in $\tau$ time) from normal variations. On the segregated events, the algorithm applies sequencing and recreates the compressor cycles of the refrigeration unit. We employ estimation algorithm to represent RU state as a function of room temperature ($\mathbf{S_{ru_r}} = EA(\mathbf{T_r})$) and learn the model parameters. One must note that \system~ estimates state vector only when RU is working fine because learning requires \emph{clean} data.

\subsubsection{Parameter Initialization}
When a store comes under monitoring, there exists no data to learn the model
parameters for the particular outlet. In such a scenario, \system~ uses the parameters of a contextually `similar' store  to monitor the refrigeration unit for leakage detection until sufficient clean data from the new outlet is available to update the model parameters. To find 'similar' outlet, we compare the average daily routine of managers in each store. We do not use room temperature directly to measure `similarity' because it is possible that RU is faulty at the time of installation. The rationale for using daily schedule is that out of all the factors, room temperature is most sensitive to managers' activities. While the efficiency of the refrigeration unit, building insulation, and other building characteristics are static and change over a long span of time (in months or years), the schedule of any manager is stochastic. Therefore, if the managers of two outlets have `similar' routine, there are high chances that the hourly temperature profile will also be `similar' for those outlets (Figure~\ref{fig:similarity_oplot}). Our ranking algorithm (Algorithm~\ref{algo:ranking}) then sorts all the available stores based on the `similarity' in the daily routine of their respective managers. We use $l2$-norm to measure the similarity. When there exists no `similar' outlet, the monitoring module (of \system) uses $10^\circ$C as a default threshold~\footnote{The refrigeration units are supposed to maintain a temperature range of $5^\circ$C-$8^\circ$C in the cold-rooms. Typically, temperature remains higher (than the limits) during the working hours, and within the limits during the non-working hours. The $10^\circ$C is the median value for both working and non-working hours.}.

\subsection{Monitor for Leakage Detection}
At the end of each hour, the tuned thermal model estimates typical temperature profile of the room ($\mathbf{\tilde{T}_r}$), for the framework to look for refrigerant leakage. However, it is infeasible to estimate the temperature profile which perfectly aligns with the actual temperature profile, and the reason lies in the control strategy embedded within the internal thermostat of a typical refrigeration unit. The inbuilt thermostat of an RU uses \emph{on} hysteresis as an upper threshold and \emph{off} hysteresis as a lower threshold. When room temperature goes beyond the \emph{on} hysteresis, the thermostat turns on the compressor, and the room temperature starts decreasing. Subsequently, when the room temperature reaches the \emph{off} hysteresis, the thermostat turns off the compressor and allows room temperature to increase up to \emph{on} hysteresis. However, in real-world, the temperature readings from the inbuilt thermostat of RU differs significantly from the deployed sensor measurements. As it is difficult for the external temperature sensor to match the temperature readings as sensed by the inbuilt sensor, it is also hard to accurately predict room temperature at every time instance. If we observe deviations in the raw temperature profiles ($\mathbf{T_r}$ and $\mathbf{\tilde{T}_r}$) for anomaly detection, the misalignments will lead to false conclusions~\footnote{We discuss misalignment in detail in Section~\ref{sec:discussion} - Discussions}.

Instead, \system~ analyses hourly mean temperature to mark if the RU in a particular hour is anomalous or not. In any hour, the thermostat tries to achieving a temperature which is the average of \emph{on} and \emph{off} hysteresis. For instance, in the current scenario, outlets use 5$^\circ$C as \emph{off} hysteresis and 8$^\circ$C as $on$ hysteresis to maintain an average temperature of 6.5$^\circ$C in the room. Though the temperature at any time instance may misalign, the average temperature in that hour should remain close to 6.5$^\circ$C. Thus, \system~ first computes mean actual ($T_r^h$) and estimated ($\tilde{T}_r^h$) temperature in Equation~\ref{eq:mean_temperature} followed by decision boundary for leakage detection in Equation~\ref{eq:boundary}.

\begin{equation}
    T_r^h = \frac{\sum_h{\mathbf{T_r}}}{(3600/\tau)}, \quad
    \tilde{T}_r^h = \frac{\sum_h{\mathbf{\tilde{T}_r}}}{(3600/\tau)}
    \label{eq:mean_temperature}
\end{equation}

\begin{equation}
    \hat{T}_r^h = \tilde{T}_r^h + \sigma_r^h
    \label{eq:boundary}
\end{equation}

\begin{figure*}
    \begin{minipage}{0.49\linewidth}
		\begin{algorithm}[H]
			\SetAlgoNoLine
			\tcc{Sensed and Estimated Room Temperature}	
			\KwIn{$T_r^h$, $\hat{T}_r^h$, $h_{mon}$, $an_{lock}$, $b_r$}
			\KwOut{$b_r$}

			\BlankLine
			\tcc{Update Bucket Value}
			
			\uIf{$T_r^h > \hat{T}_r^h$}{
    			$b_r = b_r + 1$\;
    			$an_{lock} = h_{mon}$; \tcp{Reset the lock}
			}
			\uElseIf{$T_r^h \leq \hat{T}_r^h$}{
				$an_{lock} = an_{lock} - 1$\;
				\uIf{$an_{lock} == 0$}{
					$b_r = 0$; \tcp{Reset the bucket}
				}
				\uElse{
					$b_r = b_r - 1$\;
				}
			}
      		\caption{Update Bucket}
			\label{algo:bucket}
		\end{algorithm}
    \end{minipage}
    \hspace{0.01\linewidth}
    \begin{minipage}{0.49\linewidth}
    	\includegraphics[width=\linewidth]{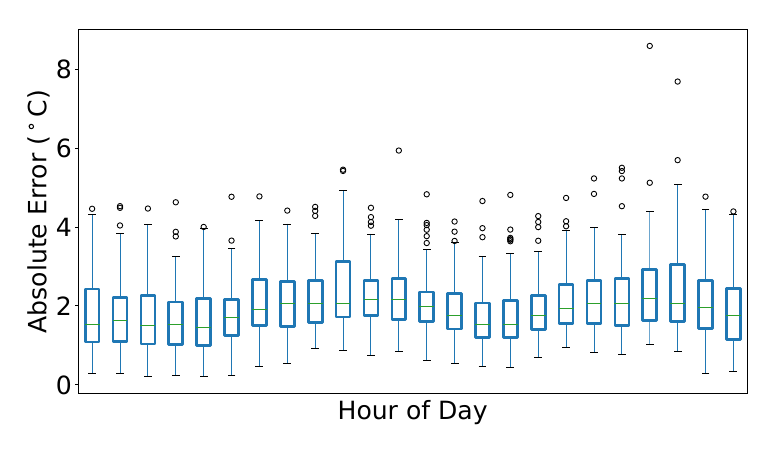}
       	\caption{During working hours, the outlet managers are much more noisy than the non-working hours.}
        \label{fig:outlet_managers}
    \end{minipage}
\end{figure*}

Here, $\tilde{T}^h_r$ is the estimated mean temperature, $\sigma^h_r$ is the standard deviation in $\mathbf{T_r}$ in an hour $h$, and $\hat{T}^h_r$ denotes the estimated decision boundary for the particular hour $h$ and room $r$. If the sensed mean temperature ($T^h_r$) is beyond the estimate ($\hat{T}^h_r$), then we mark that specific hour anomalous (Algorithm~\ref{algo:bucket}). Furthermore, to gain confidence that deviations are due to leakage, \system~ applies CUSUM (Cumulative Sum) control strategy~\cite{wikipedia2018cusum} and maintains a bucket variable ($b_r$) to monitor consecutive such anomalous instances. Whenever room temperature goes beyond the decision boundary, the monitoring module increments the $b_r$ value by one (Algorithm~\ref{algo:bucket}). For every consecutive hour, when room temperature is within the estimated limits, $b_r$ decreases by one unit. If room temperature stays below the decision boundary for consecutive $h_{mon}$ hours, \system~ assumes RU is working fine and resets the bucket. In case of missing information, $b_r$ remains unchanged.

\subsubsection{Label the Refrigeration Unit}
In the final step, \system~ labels the refrigeration unit based on its bucket value ($b_r$) in the particular hour $h$. Our analysis indicates that if the room temperature is beyond the estimated decision boundary for consecutively 36 hours, then there are maximum chances of refrigerant leakage. Though we learned the current transition thresholds from the data, we can always adjust these settings based on user requirements.

\section{Evaluation}
\label{sec:evaluation}
\begin{figure}[tp]
    \begin{minipage}[t]{0.24\linewidth}
        \centering
        \includegraphics[width=\linewidth]{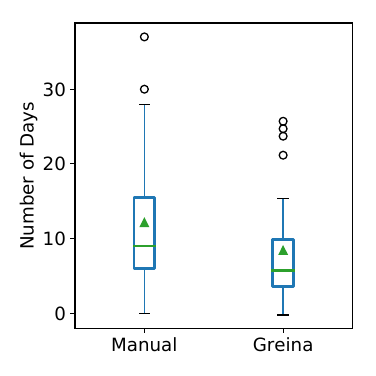}
        \subcaption{\system~ performed better or comparable to manual reporting in leakage detection.}
        \label{fig:with_manual}
    \end{minipage}
    \hspace{0.01\linewidth}
    \begin{minipage}[t]{0.36\linewidth}
        \centering
        \includegraphics[width=\linewidth]{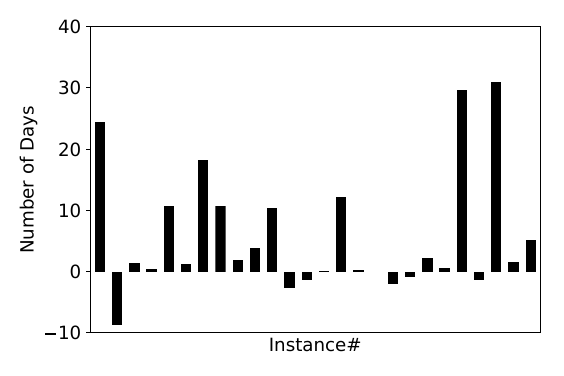}
        \subcaption{In 25 instances both \system~ and manual reported correctly identified the fault. In 19 out of 25, \system~ detected the leakage before store manager.}
        \label{fig:instance_wise}
    \end{minipage}
    \hspace{0.01\linewidth}
    \begin{minipage}[t]{0.36\linewidth}
        \centering
        \includegraphics[width=\linewidth]{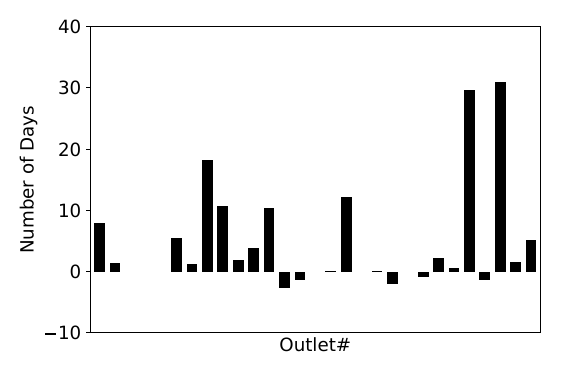}
        \subcaption{Due to delay in fixing the fault and backlog complaints, \system~ detected the fault after manual logging in six instances.}
        \label{fig:outlet_wise}
    \end{minipage}
    \caption{Our analysis indicates, a simple yet powerful framework, \system~ reduced the average leakage reporting delay by 5-6 days, significant enough to avoid energy wastage and maintain food quality by timely repairing the refrigeration unit.}
	\label{fig:comparison}
\end{figure}

For the study, we deployed a customized thermostat and power meter across 74 outlets of a retail enterprise. From July 2016 until June 2017, thermostats collected room temperature and door information from the cold-rooms, and energy meters gathered appliance level power consumption. Every cold room is $9$ ft wide, $9$ ft long, and $8$ ft high. With thick walls and doors, these highly insulated cold rooms can maintain storage temperature anywhere between $-2^\circ$C and $5^\circ$C. We keep thermostats close to the blower fan (of the RU), because that allows \system~ to focus more on the output of refrigeration unit and less on the thermal noises in the environment. We monitored weather conditions (in the region) through an API of a cloud-based weather service~\cite{wunderground}. The outlets were located in a city where outside temperature usually stays between $21^\circ$C and $37^\circ$C, all around the year. The retail enterprise maintains a log of calls from the store managers regarding the complaint in their cold-rooms. We used the fault logs for ground truth verification. During the study period, repair person identified 42 cases of refrigerant leakage across 39 outlets, in addition to several other defects. In remaining 35 stores, neither manager nor repair person mentioned any instance of refrigerant leakage in the logs during the study period. In one instance, the outlet manager called a local repair person instead of reporting the fault to the authorised maintenance contractor.

\subsection{Evaluation Metrics}
We evaluate \system~ primarily on two aspects -
\begin{enumerate*}
	\item model accuracy in estimating the decision boundary ($\mathbf{\hat{T}_{r}}$), and 
	\item minimising the delay in reporting refrigerant leakage. 
\end{enumerate*}

\subsubsection{Modelling Error}
For each hour, we compute mean absolute deviation (denoted by $e_h$) in measured ($\mathbf{T_r}$) and estimated temperature ($\mathbf{\tilde{T}_r}$) to quantify the accuracy of model parameters in simulating the room temperature in an hour $h$ (Equation~\ref{eq:error}).

\begin{equation}
	e_h = \frac{\sum_h{|\tilde{T}_r^{(t)} - T_r^{(t)}|}}{(3600/\tau)}
	\label{eq:error}
\end{equation}

\begin{equation}
    rd_m = dt_m - dt_s \quad
    rd_g = dt_g - dt_s
    \label{eq:delay}
\end{equation}

\subsubsection{Delay in Reporting the Leakage}
We labelled all the leakage instances by start date ($dt_s$) - when symptoms
became visible in the data, and end date ($dt_e$) - when repair person repaired the leak. Reporting delay is the number of days between the start and the leakage reporting dates. Equation~\ref{eq:error} computes reporting delay for store manager ($rd_m$) and \system~($rd_g$). Here, $dt_m$ and $dt_g$ are the dates when the manager and \system~ reported the leaks, respectively.

\begin{figure}[tp]
    \begin{minipage}[t]{0.31\linewidth}
        \centering
        \includegraphics[width=\linewidth]{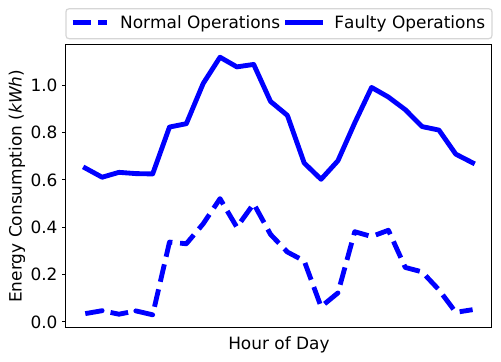}
        \subcaption{Leakage increases the power consumption of RU by 4$x$ and 2$x$ during non-working and working hours, respectively.}
        \label{fig:power_diff}
    \end{minipage}
    \hspace{0.01\linewidth}
    \begin{minipage}[t]{0.31\linewidth}
        \centering
        \includegraphics[width=\linewidth]{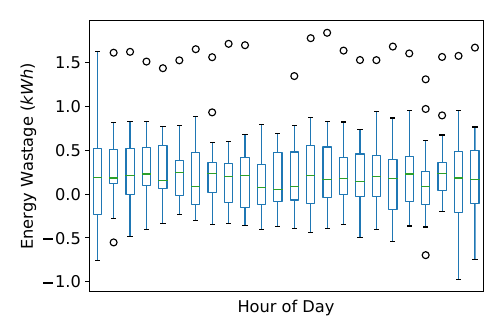}
        \subcaption{Within reporting delay period ($d_m^g$), outlets wasted significant energy in both working and non-working hours of the day.}
        \label{fig:energy_wastage}
    \end{minipage}
    \hspace{0.01\linewidth}
    \begin{minipage}[t]{0.31\linewidth}
        \centering
        \includegraphics[width=\linewidth]{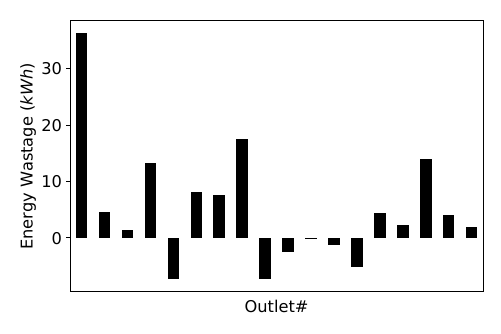}
        \subcaption{The company could have saved 5-6 $kWh$ energy every day during $d_m^g$ period which is twice the daily consumption.}
        \label{fig:minimise_wastage}
    \end{minipage}
    \begin{minipage}[t]{0.31\linewidth}
        \centering
        \includegraphics[width=\linewidth]{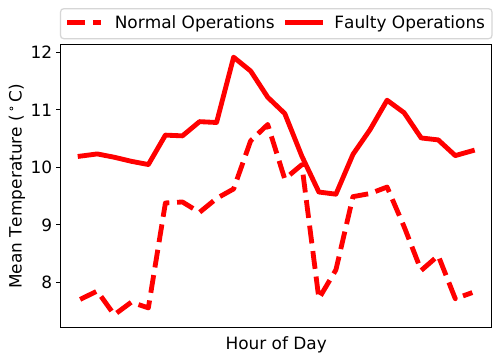}
        \subcaption{When compared with normal operations, the temperature increased by 2$^\circ$C-5$^\circ$C across the day, during refrigerant leakage.}
        \label{fig:temperature_diff}
    \end{minipage}
    \hspace{0.01\linewidth}
    \begin{minipage}[t]{0.31\linewidth}
        \centering
        \includegraphics[width=\linewidth]{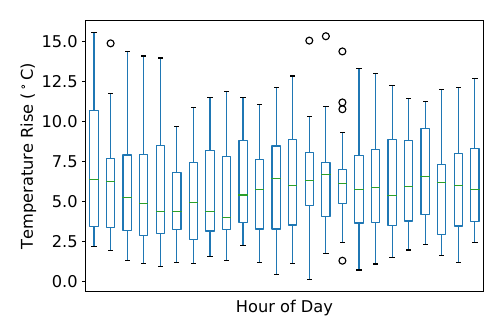}
        \subcaption{The increase in temperature is evident across all the outlets; thus, risking perishable items every day during the reporting delay.}
        \label{fig:cool_waste}
    \end{minipage}
    \hspace{0.01\linewidth}
    \begin{minipage}[t]{0.32\linewidth}
        \centering
        \includegraphics[width=\linewidth]{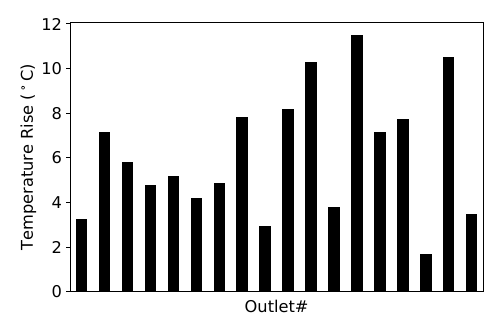}
        \subcaption{Early detection of refrigerant leakage by \system~ had kept the rooms 5$^\circ$C-6$^\circ$C colder during the reporting delay period ($d_m^g$).}
        \label{fig:minimise_cwaste}
    \end{minipage}
	\caption{Beyond being accurate in detecting the gas leakages, \system~ can also save energy and keep the room 5$^\circ$C-6$^\circ$C colder, every day during the $d_m^g$ period - the number of days between manual reporting and leakage detection by \system.}
	\label{fig:evaluation}
\end{figure}

\subsection{Model Validation}
Across all the outlets, our analysis indicates that the model tuned with \emph{clean} data (when RU was fit) can simulate the room temperature with a mean absolute error of 2$^\circ$C (with $\sigma$ = 0.9$^\circ$C), as shown in (Figure~\ref{fig:outlet_managers}). Estimation error is primarily due to misalignment and noise due to random events - such as leaving the door open, refilling the food products at a higher temperature. For the same reason, the error is usually higher during the operational hours, as also evident from the bumps. Erroneous estimations might mislead \system~ that room temperature is above the estimated temperature, but adding standard deviation in Equation~\ref{eq:boundary} minimises such instances.

\subsection{Results and Analysis}
In the study period, we noted 42 instances in the logs where either the store manager or the repair person mentioned the keywords - \emph{gas}\footnote{local people often use the word `gas' to refer refrigerant} and \emph{leakage}. In 4 cases, though the records had the \emph{gas} keyword, no leakage occurred in the refrigeration unit at that time, as validated by the store manager and the data. Furthermore, refrigerant leakage shows physical symptoms, such as water dripping and ice formation, which are visible before any leakage pattern in the temperature data. In 3 such instances, outlet managers quickly intimated the maintenance contractor to fix the refrigeration unit and \system~ had no data to analyse the symptoms for leakage detection. In remaining 35 instances, the proposed framework correctly identified 25 leaks and failed to detect leakage in the remaining scenarios. While few of them were genuine system failures, others happened due to ungovernable circumstances -
\begin{enumerate}
	\item \textbf{Too Early to Detect:} During the learning phase, \system~ borrows model parameters from a `similar' outlet. While \system~ successfully identified six leaks when knowledge transferred from one store to another, it failed twice.
	\item \textbf{Improper Learning:} Outlet managers generate significant thermal noise in the room through their dynamic and random activities. While the model can deal with such noise with a substantial amount of data, \system~ incorrectly learns the model parameters when the data is insufficient. As the occupants' behaviour differs significantly across the outlets, there exists no definite way to compute - \emph{How much data will suffice to train the model correctly?}
	\item \textbf{Noisy Sensor:} In two cases, we noticed that a faulty sensor was providing incorrect temperature readings which resulted in false negative. However, we believe that these issues are solvable with better governance around the deployment.
	\item \textbf{Low on Refrigerant:} Quite often, a refrigeration unit is actually low on refrigerant (due to heavy usage), and there exists no leakage. In such scenarios, though temperature remains in a higher range as RU is not running at the full capacity, but differs from the temperature patterns as in the case of refrigerant leakage. Therefore, when RU was low on refrigerant in 3 such instances, the technician mentioned \emph{gas top up} in the log; however, \system~ failed to find the symptoms of refrigerant leakage.
\end{enumerate}

Furthermore, \system~ reported only 6 instances of false positives - marked RU as leaking while it was working fine. In four out of six cases, either store manager or technician shut down the refrigeration unit for construction and the repairing work. As the shutdown was unexpected, \system~ confused the rise in room temperature with leakage and marked the refrigeration unit as leaking. In remaining two instances, RU broke abruptly due to an electrical fault and stayed down until a technician came to fix it. Consequently, room temperature went high and \system~ raised a refrigerant leakage flag. As the company, maintenance contractor, and the outlet managers are usually aware of these situations; we believe these flags are harmless. In addition to this, \system~ also pinpointed a case of refrigerant leakage where store manager called an unauthorised technician to repair the RU and didn't notify the enterprise. Though such cases are rare, they are a serious concern for the company.

\subsection*{Beyond Accuracy}
Though accuracy is essential, minimisation of reporting delay is a more significant concern for the stakeholders because the delay is directly proportional to energy wastage, health hazards, and product wastage. In our analysis, we noted an average reporting delay of seven days for \system, while managers had a mean reporting delay of 12 days (Figure~\ref{fig:with_manual}). In 19 out of 25 instances, \system~ detected leakage before the store manager (Figure~\ref{fig:comparison}). In eight cases, the difference was as high as 10-30 days. The early detection of leakage exhibits several benefits.

\begin{figure}[tp]
    \begin{minipage}[t]{0.32\linewidth}
        \centering
        \includegraphics[width=\linewidth]{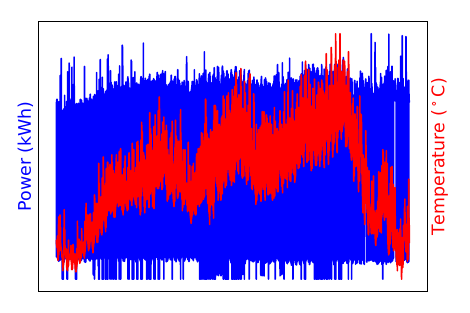}
        \subcaption{When ice forms due to other reasons (such as dirty filters), the peak power consumption remains consistent.}
        \label{fig:dirty_filter}
    \end{minipage}
    \hspace{0.01\linewidth}
    \begin{minipage}[t]{0.32\linewidth}
        \centering
        \includegraphics[width=\linewidth]{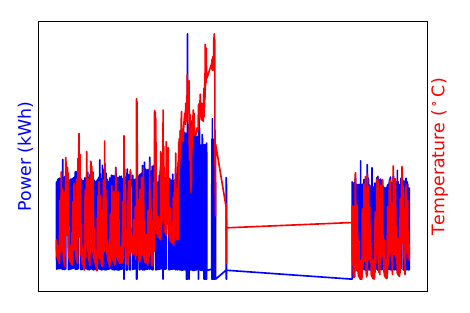}
        \subcaption{Before failure, motor within the RU draws high current which increases the peak power consumption of RU by 1.5$x$.}
        \label{fig:motor_failure}
    \end{minipage}
    \hspace{0.01\linewidth}
    \begin{minipage}[t]{0.32\linewidth}
        \centering
        \includegraphics[width=\linewidth]{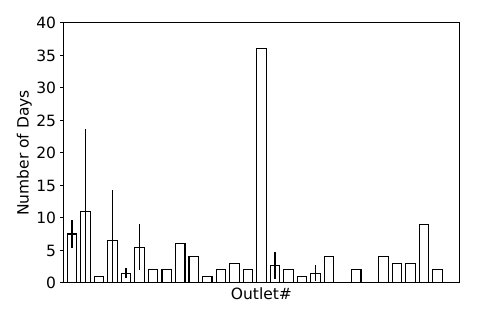}
        \subcaption{Even though outlet managers timely report the leakage, the technicians introduce a significant delay in repairing the RU.}
        \label{fig:tech_delay}
    \end{minipage}
    \caption{[Left and Middle] In addition to leakage, \system~ also reported 28 instances of ice formation and motor failure during the study. [Right] \system~ can be extended to monitor the efficiency of maintenance contractor in repairing a reported fault.}
	\label{fig:beyond_leakage}
\end{figure}

\subsubsection*{Minimise Energy Wastage}
In Section~\ref{sec:introduction}, we presented a scenario (Figure~\ref{fig:faulty_ru}) when refrigeration unit had a gas leakage, and outlet manager kept using the RU for more than a week. While such situations are common across all the stores, we noticed significant energy wastage in doing so. Figure~\ref{fig:power_diff} compares hourly energy consumption during the normal operations with faulty operations, when RU has a leakage. There are two essential takeaways -
\begin{enumerate*}
    \item energy consumption increases significantly during the working hours (almost $4x$), and 
    \item energy consumption is very high when refrigeration unit is leaking (around $7x$ during the non-working hours).
\end{enumerate*}

\begin{equation}
    d_m^g = dt_g - dt_m
    \label{eq:reporting_delay}
\end{equation}

While the activities (or routine) of store managers are hard to change, \system~ seems robust enough to minimise the energy wastage due to gas leakage. Figure~\ref{fig:energy_wastage} compares the hourly energy wasted in the $d_m^g$ period - the number of days between the reporting dates of manager and \system~(Equation~\ref{eq:reporting_delay}). Our analysis indicates that if central maintenance team had taken the recommendations from \system, they might have saved 5-10~$kWh$ energy every day (when RU was faulty) which is twice the typical daily power consumption of RU (Figure~\ref{fig:minimise_wastage}). The negative wastage depicted those scenarios when RU stopped working, and the outlet ended up consuming less energy in comparison to normal operations.

\subsubsection*{Minimising Risk to Product Quality}
Next, we observed that room temperature remains significantly high during the gas leakage (Figure~\ref{fig:temperature_diff}). High temperature risks the product quality and impacts the store operations. The average room temperature increased by 2$^\circ$C-3$^\circ$C in both working and non-working hours. When we analysed across all the 19 instances where \system~ reported before the store manager, we noted a median increase of 6$^\circ$C during the $d_t^g$ period for each hour of the day, across different outlets (Figure~\ref{fig:cool_waste}). If maintenance team had taken the recommendations from \system, they could have kept the rooms 5$^\circ$C-6$^\circ$C colder every day (when RU was faulty), by timely repairing the RU (Figure~\ref{fig:minimise_cwaste}).

Even though store managers have a benefit of observing the physical symptoms of leakage, \system~ identified 19 out of 25 leakages before the store manager. Though in two instances, \system~ couldn't detect the leakage before manager; we also noted that in remaining three cases, it occurred due to the negligence of the maintenance contractor. While outlet manager timely reported the leakage to the maintenance team, the maintenance contractor didn't take any action for several days. Consequently, the symptoms became visible and \system~ detected the leakage even though store manager repeatedly complained about the leakage.

\subsection*{Beyond Leakage}
Though we designed \system~ for refrigerant leakage, we observed that the proposed framework also identified other common time-varying faults.

\subsubsection*{Ice Formation}
Dirty air filters, defective evaporator or condenser fans, lack of refrigerant often results in the formation of ice in the refrigeration unit. The ice obstructs the path of cold air and the refrigeration unit works at a reduced efficiency. Though the symptoms of ice formation in temperature data may depend on the fault, Figure~\ref{fig:dirty_filter} depicts one scenario where the increase in temperature is akin to refrigerant leakage (Figure~\ref{fig:faulty_ru}). 
If the ice forms due to heavy usage or a dirty filter, store managers usually clean the filters, however, if the ice forms due to a fault, ice keeps on forming even after cleaning and manager needs to call the technician to repair the RU. \system~ identified 18 such instances where ice formed due to dirty air filter or defective fans.

\subsubsection*{Condenser Motor Failure}
The job of the condenser is to cool the high-pressure refrigerant gas received from the compressor by moving outside air across the condenser coils. Through condensation, the high-pressure, high-temperature refrigerant gas changes to low-temperature liquid refrigerant. However, due to wear and tear, and high temperature during the summers, the condenser fan motor usually fails and RU stops cooling the room. In an attempt to cool the room during the motor failure, RU starts drawing more current which increases the peak power consumption of RU and heats up the system further (Figure~\ref{fig:motor_failure}). Eventually, RU breaks down, and manager calls a technician to repair the motor. \system~ detected ten such instances of motor failure and raised the alarm 1-2 days before complete shutdown of the RU.

To conclude, our analysis on a reasonably rich dataset collected from actual field deployments, indicates that \system~ possesses the power to \emph{timely} identify the refrigerant leakage instances. Beyond being accurate, on an average, \system~ also proved to reduce the reporting delay by five to six days. The improvement in reporting delay can minimize the energy wastage and maintain desired temperature for the stored items. Interestingly, to achieve these benefits, the company needs to only upgrade their traditional thermostat to a smart thermostat with \system~ running in the cloud and leveraging the data collected from the thermostat. Our study indicates that with smart thermostats, the simple yet powerful framework, \system~ is easily scalable.

\section{Discussion}
\label{sec:discussion}
In this paper, we proposed a leakage detection framework \system~ for smart thermostats and validated its efficacy on a year-long field data collected from 74 outlets of a retail enterprise. In collaboration with an energy-analytic firm, we are currently deploying \system~ across all the 74 outlets. In this section, we discuss three possible dimensions to extend the proposed framework.

\subsection{Modelling Error}
We rely on our thermal model to first learn a decision boundary for \system~ to monitor the RU for any leakage on a daily basis. Though current model estimates room temperature with a mean absolute error of $2^\circ$C (with $\sigma$ = $0.9^\circ$C), the reader should also be mindful of the fact that the model is a replaceable module of the whole framework. In Appendix~\ref{app:extension}, we discuss one such extension of the current thermal model which can estimate room temperature with an RMSE of $2.85^\circ$C (with $\sigma = 1.3^\circ$C) even in a highly noisy environment of residential apartments. In our analysis, we noticed that \emph{Misalignment} and \emph{Constant Thermal Noise} are the primary sources of error. 

\begin{figure}[tp]
    \begin{minipage}[t]{0.25\linewidth}
        \centering
        \includegraphics[width=\linewidth]{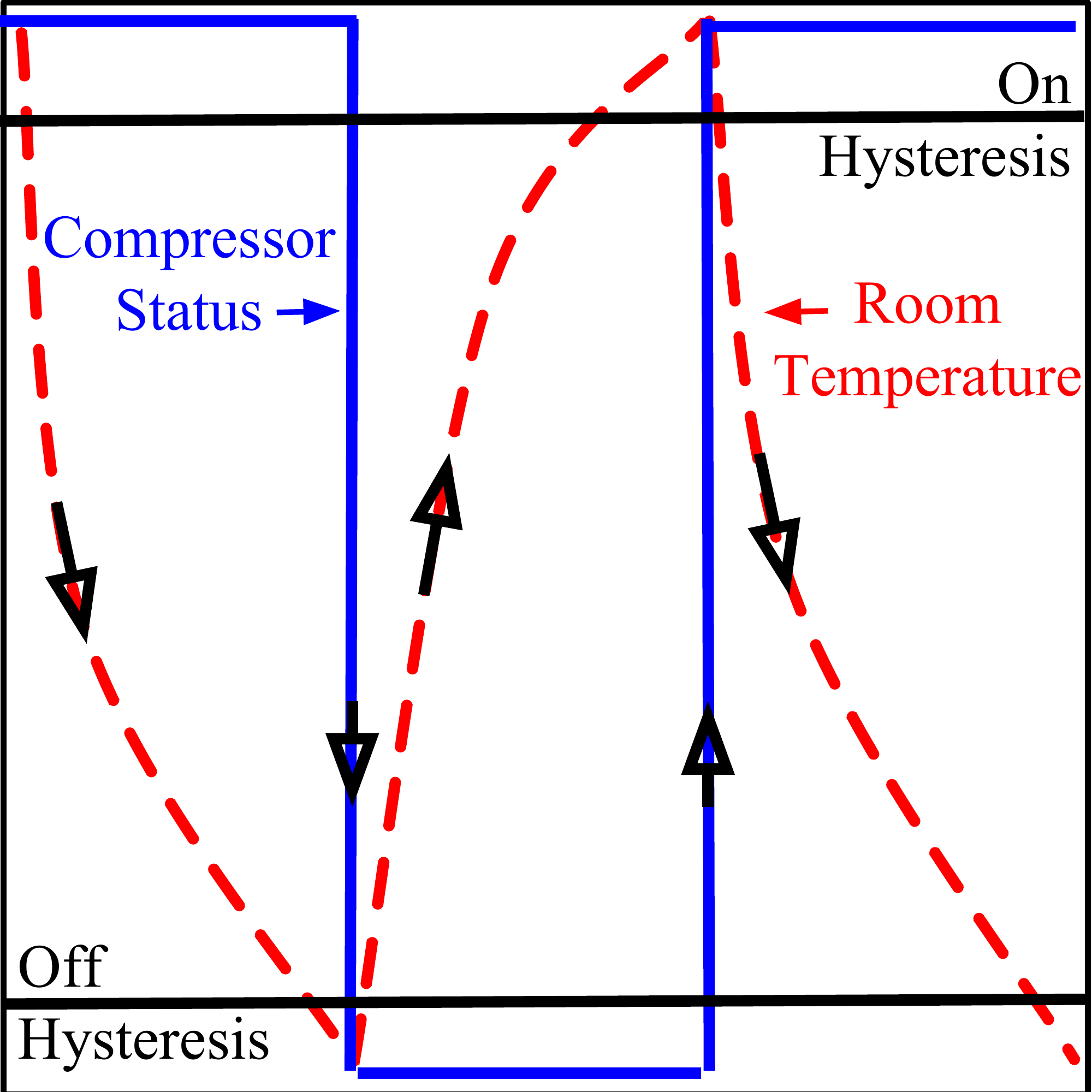}
        \subcaption{For a compressor cycle, the room temperature stays within \emph{on} and \emph{off} hysteresis.}
        \label{fig:comp_cycle}
    \end{minipage}
    \hspace{0.01\linewidth}
    \begin{minipage}[t]{0.34\linewidth}
        \centering
        \includegraphics[width=0.9\linewidth]{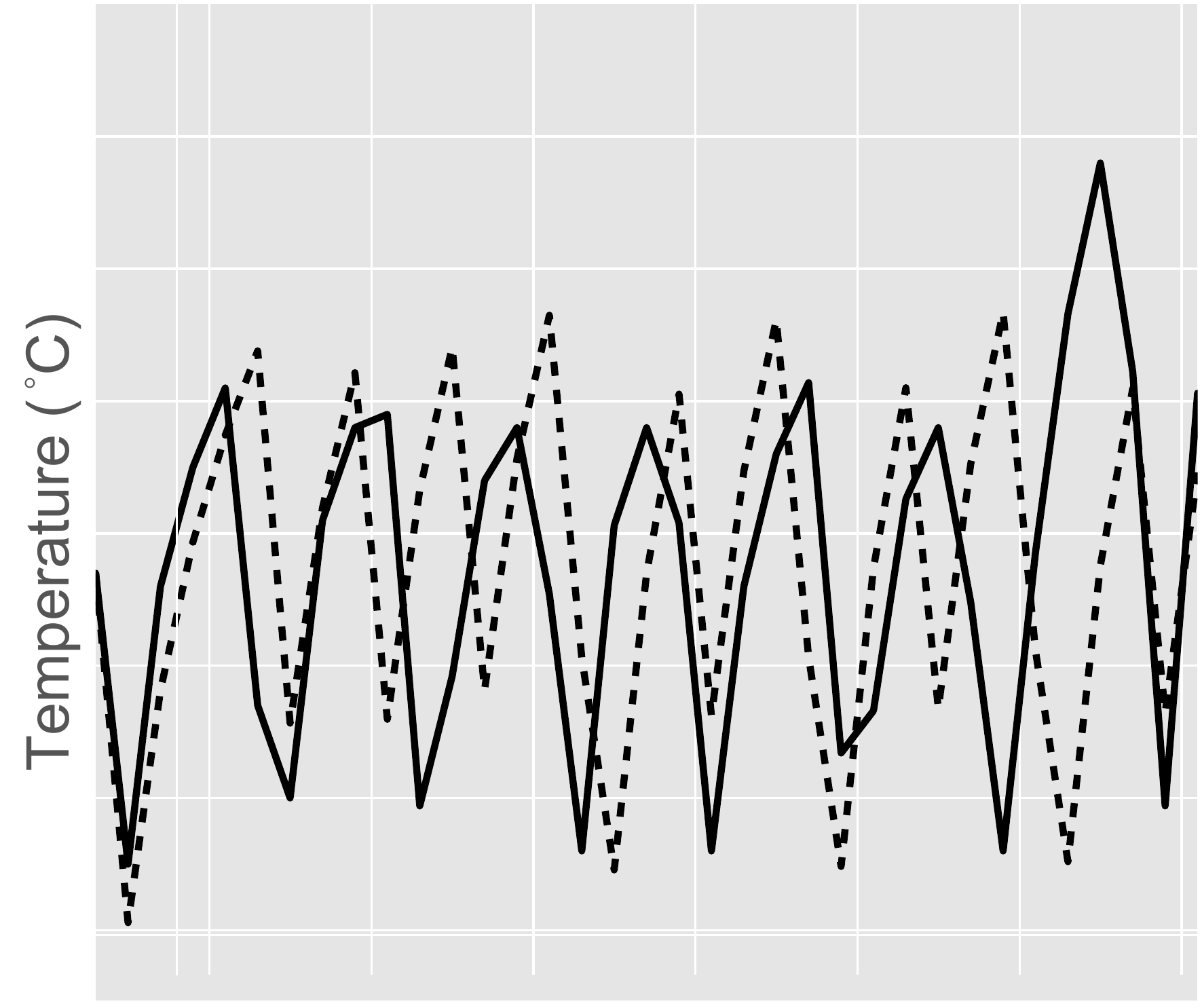}
        \subcaption{While hard line depicts recorded temperature signal, dashed line represents estimated temperature signal.}
        \label{fig:sample2}
    \end{minipage}
    \hspace{0.01\linewidth}
    \begin{minipage}[t]{0.33\linewidth}
        \centering
        \includegraphics[width=0.9\linewidth]{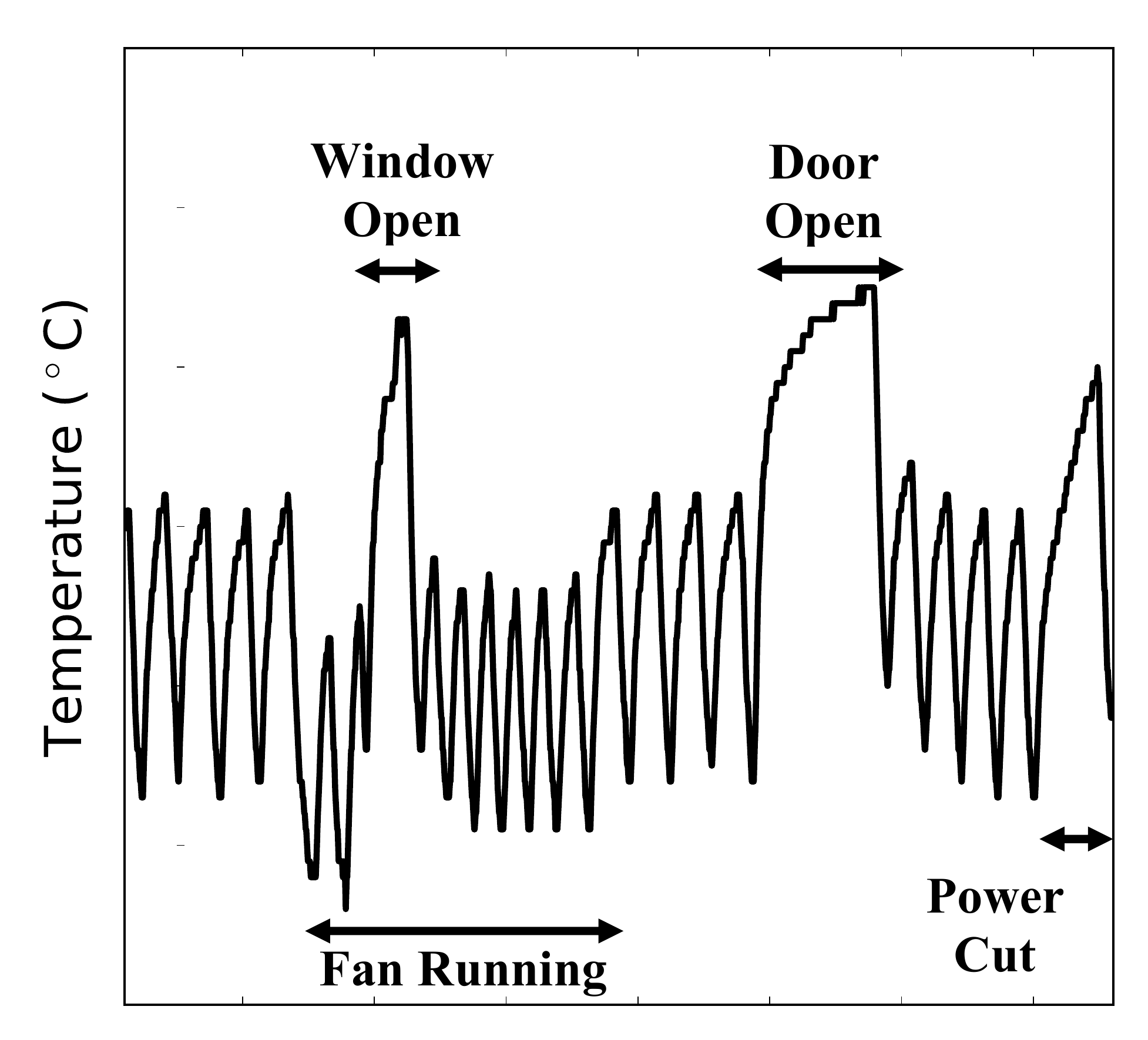}
        \subcaption{Stochastic user activities (especially in residential apartments) require time-varying noise in the thermal model.}
        \label{fig:dynamic_noise}
    \end{minipage}
    \caption{[Left and Middle] Inability to accurately estimate room temperature close to \emph{on} and \emph{off} hysteresis leads to misalignment in estimated and recorded temperature signals. [Right] Current model is unaware of dynamic user activities and assumes constant thermal noise in the room. It is another major source of modelling error, especially in residential apartments.}
	\label{fig:modelling_error}
\end{figure}

\subsubsection*{Misalignment}
When we turn on the RU (also applicable on AC), the compressor turns on and starts putting cold air into the room. Due to the flow of cold air, room temperature drops up to a certain level - \emph{off} hysteresis and compressor turns off (as depicted in Figure~\ref{fig:comp_cycle}). At this point,  RU allows room temperature to increase up to a certain level - \emph{on} hysteresis. As temperature reaches the \emph{on} hysteresis, compressor turns on again and starts cooling the space.

The control algorithm of RU decides compressor state based on the return air temperature which is measured inside the RU, near the filters. Let's call it $T_{\mathit{rat}}$. On the other hand, we install temperature sensor just outside the RU, near the fan. Now, let's say $T_r$ represents room temperature as measured by the sensor and $\Tilde{T}_r$ depicts the estimated room temperature. As per the control logic, the controller will turn off the compressor when $T_{\mathit{rat}} \geq T_{\mathit{off}}$. However, given the complex non-linear thermodynamics, it is impractical to precisely estimate $T_{\mathit{rat}}$ or $T_r$ at any time instance; thus, $T_{\mathit{rat}} - T_{\mathit{off}}$ and $T_{\mathit{rat}} - T_{\mathit{on}}$. As a result, $T_{\mathit{rat}} - T_{\mathit{off}} \neq \tilde{T}_{r} - T_{\mathit{off}}$ and $T_{\mathit{rat}} - T_{\mathit{on}} \neq \tilde{T}_{r} - T_{\mathit{on}}$. As a result, though the estimated room temperature will follow the pattern, it will fail to align perfectly with the measured room temperature. As depicted in Figure~\ref{fig:sample2} also, the actual (hard black line) and estimated (dashed black line) temperature signals are misaligned. To minimize the effect of error due to misalignment, we consider hourly mean temperature and add standard deviation to monitor the RU for refrigerant leakage.

\subsubsection*{Constant Thermal Noise} 
Another primary reason is the thermal noise due to multiple dynamic activities at the same time. In current implementation (Equation~\ref{eq:thermal_model} and Equation~\ref{eq:gmodel_wall}-\ref{eq:gmodel_lir}), we assume that the thermal noise is constant at any time. However, if the noise is coming from multiple sources, the impact of noise on room temperature will vary with time. For instance, consider Figure~\ref{fig:dynamic_noise} where initially noise is coming through two activities (window open and ceiling fan running), however, later noise is coming only through an open door. As the current model is unaware of time-varying noise, it will compute room temperature with time-independent noise learned from the data; thus inflating the error. While constant noise makes little impact on room temperature estimation in cold rooms (space is highly insulated and activities are limited), it causes a significant estimation error for residential apartments. One way to handle time-varying noise is to first identify the activity (based on temperature, occupancy, and time of the day), and then compute thermal noise for that particular time instance. At this point, we repeat that the model is a replaceable module of the whole framework; thus, the community is encouraged to explore other such variants of the current model to enhance \system's performance.

\subsection{False Negatives \& Positives}
We discussed multiple scenarios when \system~ generated both, false negatives and false positives. After scrutinizing those situations, we noted that we could avoid many of those instances through user-friendly and interactive interfaces (or advanced notification systems). Basically, by involving users, we can empower \system~ with two-way communication with the outlet managers for accurate estimation of the refrigerant leakage. In several instances, managers spotted unusual events, such as high temperature or ice formation. If they could have notified \system~ through a device, \system~ would have used the information to detect the leak,  even early. Similarly, we discussed multiple cases where store manager timely reported the leakage, but the repair person (or the maintenance contractor) only refilled the gas instead of fixing the leakage. Consequently, refrigeration unit again went down in a couple of weeks and company had to suffer from the business loss, in addition to typical consequences of refrigeration leakage. If \system~ could validate the existence of unusual pattern even after the corrective action, the maintenance company could reassess the appliance. Not only the two-way communication would minimize such instances, but it will also develop a sense of trust for the system.

However, one must also remember that user attention is costly. When users are involved, the line between a useful system and an annoying system is typically very thin. User input is valuable only if the system is designed while considering -  
\begin{enumerate}
    \item How frequently user should input the information?
    \item How much time does it take the user to input the information?
    \item How frequently system should notify the users?
    \item Is the interface intuitive and user-friendly?
\end{enumerate}

For the same reason, the new frontiers opening in the domain of smart and ambient notification and attention management systems, the design of web and mobile application, and many such factors can significantly influence the outcome of this study. In the future, we plan to implement the proposed framework (along with the user feedback) to critically evaluate the effectiveness of \system~ in achieving the desired goals at a much larger scale.

\subsection{Beyond Temperature Signal}
From data, we noticed that power consumption data provides better visibility of leakage than the room temperature because it is less sensitive to thermal noise in the environment and occupants' dynamic activities. We believe, for the same reason, technicians analyze electrical component such as input/output voltage and supply current for fault detection. Moreover, with the advent of advanced techniques to estimate usage and operating conditions of the appliance through NILM (Non-Intrusive Load Monitoring)~\cite{batra2016gemello, batra2017matrix} and EMI (Electromagnetic Interference) signatures~\cite{chen2015dose, gupta2010electrisense, Sense, gulati2014depth}, the use of energy signal for fault detection is worthwhile. However, even though the power signal is useful for fault detection, power signal alone won't allow us to monitor the consequences\footnote{When a fault occurs, the power signal cannot tell if the appliance is maintaining a suitable temperature for the occupants (if the aim is maintain user comfort), or the stored products (when using for refrigeration).}. Although, a smart thermostat together with appliance-level power consumption monitoring could provide best of both the worlds; one must also note that the duo could significantly impact the scalability of \system. Until now, \system~ only needed a smart thermostat, however, to realize the pair, the home must either have a smart plug to monitor AC power consumption, or a smart meter to get AC power consumption profile by disaggregating meter-level data through NILM algorithms.

If we first look at smart plugs, besides cost, significant variation in supply voltage, type of socket, and power quality across countries challenge the ubiquity of smart plugs. For instance, domestic appliances in India works at 230 V at 50 Hz frequency, however, in United States, 120 V power is supplied at 60 Hz for domestic use. In addition to that, voltage fluctuations, frequency variation, generation of spikes, high earth current leaking are some of the many power quality issues (especially in developing economies) that further hinders the ubiquity of smart plugs. On the other hand, if we consider smart meter data, the effectiveness of the duo would rely on the accuracy of disaggregation algorithm in capturing the AC power consumption signal. If we look at power consumption data (blue curve) in Figure~\ref{fig:faulty_ru}, refrigerant leakage mainly impacts peak power consumption (decreases with time) and the duration of each cycle (increases with time). Since peak power consumption is critical for leakage detection, the disaggregation algorithm should ensure that we correctly capture the drop in peak value, and not confuse with the power consumption of small appliances. Though there are challenges, we believe that a study exploring the effectiveness of smart thermostats along with smart meter/plug for fault detection could significantly enhance the performance of \system~ without impacting its scalability.

\subsection{Beyond Refrigeration}
Beauty of \system~ lies in its modular and systematic architecture, especially the replaceable thermal model and ability to work on top of smart thermostats. While \system~ can benefit from readily available information from the thermostat (temperature, occupancy), we can adapt it to diverse environments by tweaking the proposed thermal model. In Appendix~\ref{app:extension}, we discuss one such extension to a noisy home environment. Poor thermal insulation and time-varying stochastic activities of people are the primary two bases of thermal noise in residential apartments. In a separate analysis across five apartments for a month, we found that a non-linear thermal model can estimate room temperature with an RMSE of $2.85^\circ$C (with $\sigma = 1.3^\circ$C).

Though it is feasible to extend \system~ for the home environment, the real challenge lies in evaluating the efficacy of the proposed framework in such an environment. In residential apartments, tenants call local technicians, and typically no fault logs are available. Ground truth data collection (at a large scale) demands enormous support from tenants and the technicians in sharing the information whenever there is a fault. For the same reason, several studies in the past have either evaluated their approach in theory, or in a controlled environment. We believe a comprehensive evaluation of \system~ across residential apartments can bring up additional intriguing insights about both, the environment and the system.

\section{Conclusion}
\label{sec:conclusion}
In this paper, we discussed an unsupervised self-learning framework, \system~ that senses ambient information from smart thermostat for leakage detection. The proposed technique employs Grey-Box Modelling to estimate a decision boundary and later uses the estimates for leakage monitoring. The performance evaluation of \system~ on data from 74 stores (from a city in India) testifies that the simple yet powerful framework can reduce the reporting delay by a week with best of around 20-30 days in few instances. During these days, the retail enterprise could have saved twice the energy RU consumes on a typical day. Moreover, by timely repairing the refrigeration unit, they could have kept the rooms 5$^\circ$C-10$^\circ$C colder every day when refrigerant was leaking. In the future, we plan to deploy the retrofitted thermostats across multiple outlets and study various other challenges, as discussed in the paper.

\appendix

\begin{table*}[t]
\caption{List of symbols used in the proposed thermal model}
\centering
\ra{1.2}
\begin{tabular}{@{}lll@{}}\toprule[0.3ex]
\textbf{Symbol}       & \textbf{Description}       & \textbf{Unit}       \\
\hline
$\tau$  				&   Sampling Interval         
& s \\
$r$        				&   Thermal region $\in \{r_1, r_2, r_3\}$        
& $-$ \\
$\eta_r$         	&   Thermal noise in region $r$     
& $-$ \\
$Q_{oc}$         	&   Cooling load due to occupants and their activities
& $-$ \\
$Q_{ac}$         	&   Cooling capacity of AC
& $kW$ \\
$T_{r}^{(t)}$     	&   Temperature in region $r$ at time instance $t$
& $^\circ C$ \\
$T_{e}^{(t)}$	&   External temperature at time instance $t$
& $^\circ C$ \\
$T_{w}^{(t)}$	&   Temperature of wall (facing outside) at time instance $t$
& $^\circ C$ \\
$C_r$         			&   Thermal capacity of region $r$
& $kJ/K$ \\
$C_{w}$         	&   Thermal capacity of wall (facing outside)
& $kJ/K$ \\
$K_{w}^{r}$       &   Heat transfer coefficient between wall (facing outside) and region $r$
& $kW/K$ \\
$K_{e}^r$        &   Heat transfer coefficient between external environment and region $r$
& $kW/K$ \\
$K_{e}^{w}$ &   Heat transfer coefficient between wall (facing outside) and weather
& $kW/K$ \\
$K_{r1}^{r2}$  	&    Heat transfer coefficient between $r1$ and $r2$
& $kW/K$ \\
$K_{r2}^{r3}$  	&    Heat transfer coefficient between $r2$ and $r3$
& $kW/K$ \\
$S_{ac}^{(t)}$	&    AC compressor state ($on$/$\mathit{off}$) at time instance $t$
& $-$ \\
$S_{oc}^{(t)}$	&    State of occupants at time instance $t$
& $-$ \\
\bottomrule[0.3ex]
\end{tabular}
\label{table:params}
\end{table*}

\section{Extension to Home Environment}
\label{app:extension}
A residential apartment differs from a cold room in two major aspects:
\begin{enumerate}
    \item Poor Insulation - Typically, thermal insulation in a residential apartment is substandard as compared to a cold room. Often, heat leaks through walls, gaps around doors and windows, and multiple such sources.
    \item Noisy Occupants - In cold room, loss of cooling only happens (given superior insulation) when manager opens the door for cleaning or shifting the goods. However, significant amount of cooling is consumed by the occupants in a residential apartment.
\end{enumerate}

Thus, we need a high order thermal model to capture the non-linearity and estimate the room temperature in a residential apartment. Equation~\ref{eq:gmodel_wall}-\ref{eq:gmodel_lir} depict one such thermal model derived from Bacher et al.~\cite{bacher2011identifying}.

\begin{equation}
	\begin{split}
	\frac{(T_{w}^{(t+1)} - T_{w}^{(t)}) \times C_{w}}{\tau} = K_{e}^{w} \times (T_{e}^{(t)} - T_{w}^{(t)}) + \sum_{r = 1}^{r_l}{K_{w}^{r} \times (T_{r}^{(t)} - T_{w}^{(t)})}
	\end{split}
	\label{eq:gmodel_wall}
\end{equation}

\begin{equation}
	\begin{split}
	\frac{(T_{r_1}^{(t+1)} - T_{r_1}^{(t)}) \times C_{r_1}}{\tau} = K_{e}^{r_1} \times (T_{e}^{(t)} - T_{r_1}^{(t)}) + K_{w}^{r_1} \times (T_{w}^{(t)} - T_{r_1}^{(t)}) + K_{r_1}^{r_2} \times (T_{r_2}^{(t)} - T_{r_1}^{(t)})\\ + Q_{ac} \times S_{ac} + \eta_{r_1}
	\end{split}
	\label{eq:gmodel_hir}
\end{equation}

\begin{equation}
	\begin{split}
	\frac{(T_{r_2}^{(t+1)} - T_{r_2}^{(t)}) \times C_{r_2}}{\tau} = K_{e}^{r_2} \times (T_{e}^{(t)} - T_{r_2}^{(t)}) + K_{w}^{r_2} \times (T_{w}^{(t)} - T_{r_2}^{(t)}) + K_{r_{1}}^{r_2} \times (T_{r_{1}}^{(t)} - T_{r_2}^{(t)})\\ + K_{r_2}^{r_{3}} \times (T_{r_{3}}^{(t)} - T_{r_2}^{(t)}) + Q_{oc} \times S_{oc} + \eta_{r_2}
	\end{split}
	\label{eq:gmodel_mir}
\end{equation}
	
\begin{equation}
	\begin{split}
	\frac{(T_{r_3}^{(t+1)} - T_{r_3}^{(t)}) \times C_{r_3}}{\tau} = K_{e}^{r_3} \times (T_{e}^{(t)} - T_{r_3}^{(t)}) + K_{w}^{r_3} \times (T_{w}^{(t)} - T_{r_3}^{(t)}) + K_{r_{2}}^{r_3} \times (T_{r_{3}}^{(t)} - T_{2}^{(t)}) + \eta_{r_3}
	\end{split}
	\label{eq:gmodel_lir}
\end{equation}

Here, the first region ($r_1$) is the area in proximity of the AC, thus facing direct and the maximal impact of cold air coming from the AC. The second region, $r_2$ is the region where occupants stay, and the region receives indirect cooling from $r_1$ region where AC is present. $r_3$ indicates corner spaces in the room. The thermal model in Equation~\ref{eq:thermal_model} is a special case of above mentioned model where whole room is considered as single region and manager's activities are monitored through a door sensor.  Table~\ref{table:params} describes all the notations used in the extended thermal model.

The idea is to logically divide the room into multiple thermal regions and capture heat transfer at region level. In the model, we assume that thermostat is installed closer to the AC and each region is considered to be separated by a thin layer of air having negligible thermal mass. To evaluate the model, we installed smart thermostat in the bedrooms of five residential apartments, and collected temperature and occupancy data for a month. In parallel, we gathered weather information from a cloud based weather service. Leave p-out cross validation (with $p=10$) indicates that even in such a noisy environment, the extended model can estimate room temperature with a mean RMSE of $2.85^\circ$C (with $\sigma = 1.3^\circ$C).

\begin{equation}
	\theta = \{C_{w}, C_{r_1}, C_{r_2}, C_{r_3}, K_{w}^{r_1}, K_{w}^{r_2}, K_{w}^{r_3}, K_{e}^{r_1}, K_{e}^{r_2}, K_{e}^{r_3}, K_{e}^{w}, K_{r_1}^{r_2}, K_{r_2}^{r_3}, \eta_{r_1}, \eta_{r_2}, \eta_{r_3}, Q_{ac}, Q_{oc}\}
\label{eq:parameters}
\end{equation}

One must note that we only used temperature in $r_1$ ($\mathbf{T_{r_1}}$), outside temperature ($\mathbf{T_e}$), and occupancy information ($\mathbf{S_{oc}}$) to learn the parameter set $\theta$ (Equation~\ref{eq:parameters}). While temperature and motion data can certainly be captured from a smart thermostat, climatic conditions are readily available from cloud based weather services. Once \system~ learns the model parameters, it will simulate temperature in $r_1$ and compare with actual temperature data for leakage detection. Details for monitoring stage remain same as specified in Section~\ref{sec:approach}. We believe, an interchangeable thermal model empowers \system~ to be genuinely pervasive and ubiquitous. Depending upon the design requirements, researchers can explore different variants of the thermal model for diverse scenarios.

\bibliographystyle{ACM-Reference-Format}
\bibliography{greina_bibliography}


\begin{thebibliography}{72}


\ifx \showCODEN    \undefined \def \showCODEN     #1{\unskip}     \fi
\ifx \showDOI      \undefined \def \showDOI       #1{#1}\fi
\ifx \showISBNx    \undefined \def \showISBNx     #1{\unskip}     \fi
\ifx \showISBNxiii \undefined \def \showISBNxiii  #1{\unskip}     \fi
\ifx \showISSN     \undefined \def \showISSN      #1{\unskip}     \fi
\ifx \showLCCN     \undefined \def \showLCCN      #1{\unskip}     \fi
\ifx \shownote     \undefined \def \shownote      #1{#1}          \fi
\ifx \showarticletitle \undefined \def \showarticletitle #1{#1}   \fi
\ifx \showURL      \undefined \def \showURL       {\relax}        \fi
\providecommand\bibfield[2]{#2}
\providecommand\bibinfo[2]{#2}
\providecommand\natexlab[1]{#1}
\providecommand\showeprint[2][]{arXiv:#2}

\bibitem[\protect\citeauthoryear{Assawamartbunlue and
  Brandemuehl}{Assawamartbunlue and Brandemuehl}{2006}]%
        {assawamartbunlue2006refrigerant}
\bibfield{author}{\bibinfo{person}{Kriengkrai Assawamartbunlue} {and}
  \bibinfo{person}{Michael~J Brandemuehl}.} \bibinfo{year}{2006}\natexlab{}.
\newblock \showarticletitle{Refrigerant leakage detection and diagnosis for a
  distributed refrigeration system}.
\newblock \bibinfo{journal}{\emph{HVAC\&R Research}}  \bibinfo{volume}{12}
  (\bibinfo{year}{2006}), \bibinfo{pages}{389--405}.
\newblock


\bibitem[\protect\citeauthoryear{Bacher and Madsen}{Bacher and Madsen}{2011}]%
        {bacher2011identifying}
\bibfield{author}{\bibinfo{person}{Peder Bacher} {and} \bibinfo{person}{Henrik
  Madsen}.} \bibinfo{year}{2011}\natexlab{}.
\newblock \showarticletitle{Identifying suitable models for the heat dynamics
  of buildings}.
\newblock \bibinfo{journal}{\emph{Energy and Buildings}}  \bibinfo{volume}{43}
  (\bibinfo{year}{2011}), \bibinfo{pages}{1511--1522}.
\newblock


\bibitem[\protect\citeauthoryear{Batra, Singh, and Whitehouse}{Batra
  et~al\mbox{.}}{2016}]%
        {batra2016gemello}
\bibfield{author}{\bibinfo{person}{Nipun Batra}, \bibinfo{person}{Amarjeet
  Singh}, {and} \bibinfo{person}{Kamin Whitehouse}.}
  \bibinfo{year}{2016}\natexlab{}.
\newblock \showarticletitle{Gemello: Creating a Detailed Energy Breakdown from
  Just the Monthly Electricity Bill}. In \bibinfo{booktitle}{\emph{Proceedings
  of the 22Nd ACM SIGKDD International Conference on Knowledge Discovery and
  Data Mining}}. \bibinfo{publisher}{ACM}, \bibinfo{address}{New York, NY,
  USA}, \bibinfo{pages}{431--440}.
\newblock


\bibitem[\protect\citeauthoryear{Batra, Wang, Singh, and Whitehouse}{Batra
  et~al\mbox{.}}{2017}]%
        {batra2017matrix}
\bibfield{author}{\bibinfo{person}{Nipun Batra}, \bibinfo{person}{Hongning
  Wang}, \bibinfo{person}{Amarjeet Singh}, {and} \bibinfo{person}{Kamin
  Whitehouse}.} \bibinfo{year}{2017}\natexlab{}.
\newblock \showarticletitle{Matrix Factorisation for Scalable Energy
  Breakdown}. In \bibinfo{booktitle}{\emph{AAAI Conference on Artificial
  Intelligence}}. \bibinfo{publisher}{AAAI Press}, \bibinfo{address}{Palo Alto,
  CA, USA}, \bibinfo{pages}{4467--4473}.
\newblock


\bibitem[\protect\citeauthoryear{Behfar, Yuill, and Yu}{Behfar
  et~al\mbox{.}}{2017}]%
        {behfar2017automated}
\bibfield{author}{\bibinfo{person}{Alireza Behfar}, \bibinfo{person}{David
  Yuill}, {and} \bibinfo{person}{Yuebin Yu}.} \bibinfo{year}{2017}\natexlab{}.
\newblock \showarticletitle{Automated fault detection and diagnosis methods for
  supermarket equipment (RP-1615)}.
\newblock \bibinfo{journal}{\emph{Science and Technology for the Built
  Environment}}  \bibinfo{volume}{23} (\bibinfo{year}{2017}),
  \bibinfo{pages}{1253--1266}.
\newblock


\bibitem[\protect\citeauthoryear{Bender, Skrypnik, Voigt, Marcoll, and
  Rapp}{Bender et~al\mbox{.}}{2003}]%
        {bender2003selective}
\bibfield{author}{\bibinfo{person}{Florian Bender}, \bibinfo{person}{Aleksandr
  Skrypnik}, \bibinfo{person}{Achim Voigt}, \bibinfo{person}{Joachim Marcoll},
  {and} \bibinfo{person}{Michael Rapp}.} \bibinfo{year}{2003}\natexlab{}.
\newblock \showarticletitle{Selective detection of HFC and HCFC refrigerants
  using a surface acoustic wave sensor system}.
\newblock \bibinfo{journal}{\emph{Analytical Chemistry}}  \bibinfo{volume}{75}
  (\bibinfo{year}{2003}), \bibinfo{pages}{5262--5266}.
\newblock


\bibitem[\protect\citeauthoryear{Brambley, Haves, McDonald, Torcellini, Hansen,
  Holmberg, and Roth}{Brambley et~al\mbox{.}}{2005}]%
        {brambley2005advanced}
\bibfield{author}{\bibinfo{person}{Michael~R Brambley}, \bibinfo{person}{Philip
  Haves}, \bibinfo{person}{Sean~C McDonald}, \bibinfo{person}{Paul Torcellini},
  \bibinfo{person}{D Hansen}, \bibinfo{person}{DR Holmberg}, {and}
  \bibinfo{person}{KW Roth}.} \bibinfo{year}{2005}\natexlab{}.
\newblock \bibinfo{booktitle}{\emph{Advanced sensors and controls for building
  applications: Market assessment and potential R\&D pathways}}.
\newblock \bibinfo{type}{{T}echnical {R}eport}. \bibinfo{institution}{EERE
  Publication and Product Library, Washington, DC (United States)}.
\newblock


\bibitem[\protect\citeauthoryear{Breuker and Braun}{Breuker and Braun}{1998}]%
        {breuker1998common}
\bibfield{author}{\bibinfo{person}{Mark~S Breuker} {and}
  \bibinfo{person}{James~E Braun}.} \bibinfo{year}{1998}\natexlab{}.
\newblock \showarticletitle{Common faults and their impacts for rooftop air
  conditioners}.
\newblock \bibinfo{journal}{\emph{HVAC\&R Research}}  \bibinfo{volume}{4}
  (\bibinfo{year}{1998}), \bibinfo{pages}{303--318}.
\newblock


\bibitem[\protect\citeauthoryear{Chandola, Banerjee, and Kumar}{Chandola
  et~al\mbox{.}}{2009}]%
        {chandola2009anomaly}
\bibfield{author}{\bibinfo{person}{Varun Chandola}, \bibinfo{person}{Arindam
  Banerjee}, {and} \bibinfo{person}{Vipin Kumar}.}
  \bibinfo{year}{2009}\natexlab{}.
\newblock \showarticletitle{Anomaly detection: A survey}.
\newblock \bibinfo{journal}{\emph{ACM computing surveys (CSUR)}}
  \bibinfo{volume}{41} (\bibinfo{year}{2009}), \bibinfo{pages}{15}.
\newblock


\bibitem[\protect\citeauthoryear{Chen and Braun}{Chen and Braun}{2001}]%
        {chen2001simple}
\bibfield{author}{\bibinfo{person}{Bin Chen} {and} \bibinfo{person}{James~E
  Braun}.} \bibinfo{year}{2001}\natexlab{}.
\newblock \showarticletitle{Simple rule-based methods for fault detection and
  diagnostics applied to packaged air conditioners/Discussion}.
\newblock \bibinfo{journal}{\emph{ASHRAE Transactions}}  \bibinfo{volume}{107}
  (\bibinfo{year}{2001}), \bibinfo{pages}{847}.
\newblock


\bibitem[\protect\citeauthoryear{Chen, Gupta, Larson, and Patel}{Chen
  et~al\mbox{.}}{2015}]%
        {chen2015dose}
\bibfield{author}{\bibinfo{person}{Ke-Yu Chen}, \bibinfo{person}{Sidhant
  Gupta}, \bibinfo{person}{Eric~C Larson}, {and} \bibinfo{person}{Shwetak
  Patel}.} \bibinfo{year}{2015}\natexlab{}.
\newblock \showarticletitle{Dose: Detecting user-driven operating states of
  electronic devices from a single sensing point}. In
  \bibinfo{booktitle}{\emph{Pervasive Computing and Communications (PerCom),
  2015 IEEE International Conference on}}. \bibinfo{publisher}{IEEE},
  \bibinfo{address}{Piscataway, NJ, USA}, \bibinfo{pages}{46--54}.
\newblock


\bibitem[\protect\citeauthoryear{Crawley, Lawrie, Pedersen, and
  Winkelmann}{Crawley et~al\mbox{.}}{2000}]%
        {crawley2000energy}
\bibfield{author}{\bibinfo{person}{Drury~B Crawley}, \bibinfo{person}{Linda~K
  Lawrie}, \bibinfo{person}{Curtis~O Pedersen}, {and}
  \bibinfo{person}{Frederick~C Winkelmann}.} \bibinfo{year}{2000}\natexlab{}.
\newblock \showarticletitle{Energy plus: energy simulation program}.
\newblock \bibinfo{journal}{\emph{ASHRAE journal}}  \bibinfo{volume}{42}
  (\bibinfo{year}{2000}), \bibinfo{pages}{49--56}.
\newblock


\bibitem[\protect\citeauthoryear{Davis and Gertler}{Davis and Gertler}{2015}]%
        {davis2015contribution}
\bibfield{author}{\bibinfo{person}{Lucas~W Davis} {and} \bibinfo{person}{Paul~J
  Gertler}.} \bibinfo{year}{2015}\natexlab{}.
\newblock \showarticletitle{Contribution of air conditioning adoption to future
  energy use under global warming}.
\newblock \bibinfo{journal}{\emph{Proceedings of the National Academy of
  Sciences}}  \bibinfo{volume}{112} (\bibinfo{year}{2015}),
  \bibinfo{pages}{5962--5967}.
\newblock


\bibitem[\protect\citeauthoryear{Dewson, Day, and Irving}{Dewson
  et~al\mbox{.}}{1993}]%
        {dewson1993least}
\bibfield{author}{\bibinfo{person}{T Dewson}, \bibinfo{person}{B Day}, {and}
  \bibinfo{person}{AD Irving}.} \bibinfo{year}{1993}\natexlab{}.
\newblock \showarticletitle{Least squares parameter estimation of a reduced
  order thermal model of an experimental building}.
\newblock \bibinfo{journal}{\emph{Building and Environment}}
  \bibinfo{volume}{28} (\bibinfo{year}{1993}), \bibinfo{pages}{127--137}.
\newblock


\bibitem[\protect\citeauthoryear{Diagnostics}{Diagnostics}{2005}]%
        {diagnostics2005advanced}
\bibfield{author}{\bibinfo{person}{Temperature-Only Refrigeration~Cycle
  Diagnostics}.} \bibinfo{year}{2005}\natexlab{}.
\newblock \showarticletitle{Advanced Automated HVAC Fault Detection and
  Diagnostics Commercialization Program}.
\newblock \bibinfo{journal}{\emph{Contract}}  \bibinfo{volume}{500}
  (\bibinfo{year}{2005}), \bibinfo{pages}{03--030}.
\newblock


\bibitem[\protect\citeauthoryear{Dong, Gorbounov, Yuan, Wu, Srivastav, Bailey,
  and O’Neill}{Dong et~al\mbox{.}}{2013}]%
        {dong2013integrated}
\bibfield{author}{\bibinfo{person}{Bing Dong}, \bibinfo{person}{Mikhail
  Gorbounov}, \bibinfo{person}{Shui Yuan}, \bibinfo{person}{Tiejun Wu},
  \bibinfo{person}{Abhishek Srivastav}, \bibinfo{person}{Trevor Bailey}, {and}
  \bibinfo{person}{Zheng O’Neill}.} \bibinfo{year}{2013}\natexlab{}.
\newblock \showarticletitle{Integrated energy performance modeling for a retail
  store building}. In \bibinfo{booktitle}{\emph{Building simulation}},
  Vol.~\bibinfo{volume}{6}. \bibinfo{publisher}{Springer},
  \bibinfo{address}{Salmon Tower Building, NY, USA}, \bibinfo{pages}{283--295}.
\newblock


\bibitem[\protect\citeauthoryear{Downey and Proctor}{Downey and
  Proctor}{2002}]%
        {downey2002can}
\bibfield{author}{\bibinfo{person}{Tom Downey} {and} \bibinfo{person}{John
  Proctor}.} \bibinfo{year}{2002}\natexlab{}.
\newblock \showarticletitle{What can 13,000 air conditioners tell us?}
\newblock \bibinfo{journal}{\emph{the Proceedings of the 2002 ACEEE Summer
  Study on Energy Efficiency in Buildings}}  \bibinfo{volume}{1}
  (\bibinfo{year}{2002}), \bibinfo{pages}{53--67}.
\newblock


\bibitem[\protect\citeauthoryear{Du, Jin, and Wu}{Du et~al\mbox{.}}{2007}]%
        {du2007fault}
\bibfield{author}{\bibinfo{person}{Zhimin Du}, \bibinfo{person}{Xinqiao Jin},
  {and} \bibinfo{person}{Lizhou Wu}.} \bibinfo{year}{2007}\natexlab{}.
\newblock \showarticletitle{Fault detection and diagnosis based on improved PCA
  with JAA method in VAV systems}.
\newblock \bibinfo{journal}{\emph{Building and Environment}}
  \bibinfo{volume}{42} (\bibinfo{year}{2007}), \bibinfo{pages}{3221--3232}.
\newblock


\bibitem[\protect\citeauthoryear{Dube}{Dube}{2009}]%
        {dube2009refrigerant}
\bibfield{author}{\bibinfo{person}{Serge Dube}.}
  \bibinfo{year}{2009}\natexlab{}.
\newblock \bibinfo{title}{Refrigerant leak-detection systems}.
\newblock
\newblock
\newblock
\shownote{US Patent App. 12/256,504.}


\bibitem[\protect\citeauthoryear{Fisera and Hrncar}{Fisera and Hrncar}{2012}]%
        {fisera2012system}
\bibfield{author}{\bibinfo{person}{Radek Fisera} {and} \bibinfo{person}{Martin
  Hrncar}.} \bibinfo{year}{2012}\natexlab{}.
\newblock \bibinfo{title}{System and method for performance monitoring of
  commercial refrigeration}.
\newblock
\newblock
\newblock
\shownote{US Patent App. 12/976,088.}


\bibitem[\protect\citeauthoryear{Fisera and Stluka}{Fisera and Stluka}{2012}]%
        {fisera2012performance}
\bibfield{author}{\bibinfo{person}{Radek Fisera} {and} \bibinfo{person}{Petr
  Stluka}.} \bibinfo{year}{2012}\natexlab{}.
\newblock \showarticletitle{Performance monitoring of the refrigeration system
  with minimum set of sensors}.
\newblock \bibinfo{journal}{\emph{World Academy of Science, Engineering and
  Technology}}  \bibinfo{volume}{6} (\bibinfo{year}{2012}),
  \bibinfo{pages}{396--401}.
\newblock


\bibitem[\protect\citeauthoryear{Francis, Maidment, and Davies}{Francis
  et~al\mbox{.}}{2017}]%
        {francis2017investigation}
\bibfield{author}{\bibinfo{person}{Christina Francis}, \bibinfo{person}{Graeme
  Maidment}, {and} \bibinfo{person}{Gareth Davies}.}
  \bibinfo{year}{2017}\natexlab{}.
\newblock \showarticletitle{An investigation of refrigerant leakage in
  commercial refrigeration}.
\newblock \bibinfo{journal}{\emph{International Journal of Refrigeration}}
  \bibinfo{volume}{74} (\bibinfo{year}{2017}), \bibinfo{pages}{12--21}.
\newblock


\bibitem[\protect\citeauthoryear{Ganu, Rahayu, Seetharam, Kunnath, Kumar, Arya,
  Husain, and Kalyanaraman}{Ganu et~al\mbox{.}}{2014}]%
        {ganu2014socketwatch}
\bibfield{author}{\bibinfo{person}{Tanuja Ganu}, \bibinfo{person}{Dwi Rahayu},
  \bibinfo{person}{Deva~P Seetharam}, \bibinfo{person}{Rajesh Kunnath},
  \bibinfo{person}{Ashok~Pon Kumar}, \bibinfo{person}{Vijay Arya},
  \bibinfo{person}{Saiful~A Husain}, {and} \bibinfo{person}{Shivkumar
  Kalyanaraman}.} \bibinfo{year}{2014}\natexlab{}.
\newblock \showarticletitle{SocketWatch: an autonomous appliance monitoring
  system}. In \bibinfo{booktitle}{\emph{Pervasive Computing and Communications
  (PerCom), 2014 IEEE International Conference on}}. \bibinfo{publisher}{IEEE},
  \bibinfo{address}{Piscataway, NJ, USA}, \bibinfo{pages}{38--43}.
\newblock


\bibitem[\protect\citeauthoryear{Gertler}{Gertler}{2013}]%
        {gertler2013fault}
\bibfield{author}{\bibinfo{person}{Janos Gertler}.}
  \bibinfo{year}{2013}\natexlab{}.
\newblock \bibinfo{booktitle}{\emph{Fault detection and diagnosis}}.
\newblock \bibinfo{publisher}{Springer}, \bibinfo{address}{Salmon Tower
  Building, NY, USA}.
\newblock


\bibitem[\protect\citeauthoryear{Grace, Datta, and Tassou}{Grace
  et~al\mbox{.}}{2005}]%
        {grace2005sensitivity}
\bibfield{author}{\bibinfo{person}{IN Grace}, \bibinfo{person}{Datta Datta},
  {and} \bibinfo{person}{SA Tassou}.} \bibinfo{year}{2005}\natexlab{}.
\newblock \showarticletitle{Sensitivity of refrigeration system performance to
  charge levels and parameters for on-line leak detection}.
\newblock \bibinfo{journal}{\emph{Applied Thermal Engineering}}
  \bibinfo{volume}{25} (\bibinfo{year}{2005}), \bibinfo{pages}{557--566}.
\newblock


\bibitem[\protect\citeauthoryear{Gulati, Ram, and Singh}{Gulati
  et~al\mbox{.}}{2014}]%
        {gulati2014depth}
\bibfield{author}{\bibinfo{person}{Manoj Gulati},
  \bibinfo{person}{Shobha~Sundar Ram}, {and} \bibinfo{person}{Amarjeet Singh}.}
  \bibinfo{year}{2014}\natexlab{}.
\newblock \showarticletitle{An in Depth Study into Using EMI Signatures for
  Appliance Identification}. In \bibinfo{booktitle}{\emph{Proceedings of the
  1st ACM Conference on Embedded Systems for Energy-Efficient Buildings}}.
  \bibinfo{publisher}{ACM}, \bibinfo{address}{New York, NY, USA},
  \bibinfo{pages}{70--79}.
\newblock


\bibitem[\protect\citeauthoryear{Gupta, Reynolds, and Patel}{Gupta
  et~al\mbox{.}}{2010}]%
        {gupta2010electrisense}
\bibfield{author}{\bibinfo{person}{Sidhant Gupta}, \bibinfo{person}{Matthew~S.
  Reynolds}, {and} \bibinfo{person}{Shwetak~N. Patel}.}
  \bibinfo{year}{2010}\natexlab{}.
\newblock \showarticletitle{ElectriSense: Single-point Sensing Using EMI for
  Electrical Event Detection and Classification in the Home}. In
  \bibinfo{booktitle}{\emph{Proceedings of the 12th ACM International
  Conference on Ubiquitous Computing}}. \bibinfo{publisher}{ACM},
  \bibinfo{address}{New York, NY, USA}, \bibinfo{pages}{139--148}.
\newblock


\bibitem[\protect\citeauthoryear{Han, Cao, Gu, and Ren}{Han
  et~al\mbox{.}}{2010}]%
        {han2010pca}
\bibfield{author}{\bibinfo{person}{Hua Han}, \bibinfo{person}{Zhikun Cao},
  \bibinfo{person}{Bo Gu}, {and} \bibinfo{person}{Neng Ren}.}
  \bibinfo{year}{2010}\natexlab{}.
\newblock \showarticletitle{PCA-SVM-based automated fault detection and
  diagnosis (AFDD) for vapor-compression refrigeration systems}.
\newblock \bibinfo{journal}{\emph{HVAC\&R Research}}  \bibinfo{volume}{16}
  (\bibinfo{year}{2010}), \bibinfo{pages}{295--313}.
\newblock


\bibitem[\protect\citeauthoryear{Isermann}{Isermann}{2011}]%
        {isermann2011fault}
\bibfield{author}{\bibinfo{person}{Rolf Isermann}.}
  \bibinfo{year}{2011}\natexlab{}.
\newblock \bibinfo{booktitle}{\emph{Fault-Diagnosis Applications: Model-Based
  Condition Monitoring Actuators, Drives, Machinery, Plants, Sensors, and
  Fault-tolerant Systems} (\bibinfo{edition}{1st} ed.)}.
\newblock \bibinfo{publisher}{Springer Publishing Company, Incorporated},
  \bibinfo{address}{Salmon Tower Building, NY, USA}.
\newblock
\showISBNx{3642127665, 9783642127663}


\bibitem[\protect\citeauthoryear{Jain, Singh, and Chandan}{Jain
  et~al\mbox{.}}{2016}]%
        {jain2016non}
\bibfield{author}{\bibinfo{person}{Milan Jain}, \bibinfo{person}{Amarjeet
  Singh}, {and} \bibinfo{person}{Vikas Chandan}.}
  \bibinfo{year}{2016}\natexlab{}.
\newblock \showarticletitle{Non-intrusive estimation and prediction of
  residential ac energy consumption}. In \bibinfo{booktitle}{\emph{Pervasive
  Computing and Communications (PerCom), 2016 IEEE International Conference
  on}}. \bibinfo{publisher}{IEEE}, \bibinfo{address}{Piscataway, NJ, USA},
  \bibinfo{pages}{1--9}.
\newblock


\bibitem[\protect\citeauthoryear{Jeffers, McLeroy, and Sacerio}{Jeffers
  et~al\mbox{.}}{1984}]%
        {jeffers1984halogen}
\bibfield{author}{\bibinfo{person}{Edward~A Jeffers}, \bibinfo{person}{R~Philip
  McLeroy}, {and} \bibinfo{person}{Jose~L Sacerio}.}
  \bibinfo{year}{1984}\natexlab{}.
\newblock \bibinfo{title}{Halogen gas leak detector}.
\newblock
\newblock
\newblock
\shownote{US Patent 4,488,118.}


\bibitem[\protect\citeauthoryear{Jeong}{Jeong}{2005}]%
        {jeong2005refrigerant}
\bibfield{author}{\bibinfo{person}{Jin-Ho Jeong}.}
  \bibinfo{year}{2005}\natexlab{}.
\newblock \bibinfo{title}{Refrigerant leakage sensing system and method}.
\newblock
\newblock
\newblock
\shownote{US Patent App. 10/938,647.}


\bibitem[\protect\citeauthoryear{Kadle and Ghodbane}{Kadle and
  Ghodbane}{2014}]%
        {kadle2014refrigerant}
\bibfield{author}{\bibinfo{person}{Prasad~S Kadle} {and}
  \bibinfo{person}{Mahmoud Ghodbane}.} \bibinfo{year}{2014}\natexlab{}.
\newblock \bibinfo{title}{Refrigerant leak detection system}.
\newblock
\newblock
\newblock
\shownote{US Patent 8,695,404.}


\bibitem[\protect\citeauthoryear{Keres, Gomes, and Litch}{Keres
  et~al\mbox{.}}{2016}]%
        {keres2016fault}
\bibfield{author}{\bibinfo{person}{Stephen~L Keres},
  \bibinfo{person}{Alberto~Regio Gomes}, {and} \bibinfo{person}{Andrew~D
  Litch}.} \bibinfo{year}{2016}\natexlab{}.
\newblock \bibinfo{title}{Fault detection and diagnosis for refrigerator from
  compressor sensor}.
\newblock
\newblock
\newblock
\shownote{US Patent 9,513,043.}


\bibitem[\protect\citeauthoryear{Kim, Yoon, Domanski, and Payne}{Kim
  et~al\mbox{.}}{2008}]%
        {kim2008design}
\bibfield{author}{\bibinfo{person}{Minsung Kim}, \bibinfo{person}{Seok~Ho
  Yoon}, \bibinfo{person}{Piotr~A Domanski}, {and} \bibinfo{person}{W~Vance
  Payne}.} \bibinfo{year}{2008}\natexlab{}.
\newblock \showarticletitle{Design of a steady-state detector for fault
  detection and diagnosis of a residential air conditioner}.
\newblock \bibinfo{journal}{\emph{International journal of refrigeration}}
  \bibinfo{volume}{31} (\bibinfo{year}{2008}), \bibinfo{pages}{790--799}.
\newblock


\bibitem[\protect\citeauthoryear{Kim, Moon, and Yoon}{Kim
  et~al\mbox{.}}{2017}]%
        {kim2017improved}
\bibfield{author}{\bibinfo{person}{Sun~Ho Kim}, \bibinfo{person}{Hyeun~Jun
  Moon}, {and} \bibinfo{person}{Young~Ran Yoon}.}
  \bibinfo{year}{2017}\natexlab{}.
\newblock \showarticletitle{Improved occupancy detection accuracy using PIR and
  door sensors for a smart thermostat}.
\newblock \bibinfo{journal}{\emph{Building Simulation 2017}}
  \bibinfo{volume}{15} (\bibinfo{year}{2017}), \bibinfo{pages}{2753--2758}.
\newblock


\bibitem[\protect\citeauthoryear{Kim and Katipamula}{Kim and
  Katipamula}{2018}]%
        {kim2018review}
\bibfield{author}{\bibinfo{person}{Woohyun Kim} {and} \bibinfo{person}{Srinivas
  Katipamula}.} \bibinfo{year}{2018}\natexlab{}.
\newblock \showarticletitle{A review of fault detection and diagnostics methods
  for building systems}.
\newblock \bibinfo{journal}{\emph{Science and Technology for the Built
  Environment}}  \bibinfo{volume}{24} (\bibinfo{year}{2018}),
  \bibinfo{pages}{3--21}.
\newblock


\bibitem[\protect\citeauthoryear{{Knud Lasse Lueth}}{{Knud Lasse
  Lueth}}{2016}]%
        {leuth2018smart}
\bibfield{author}{\bibinfo{person}{{Knud Lasse Lueth}}.}
  \bibinfo{year}{2016}\natexlab{}.
\newblock \bibinfo{title}{Smart Thermostat market for the period 2015 to 2021}.
\newblock
\newblock
\urldef\tempurl%
\url{https://goo.gl/uRccu4}
\showURL{%
\tempurl}
\newblock
\shownote{[Online; accessed 12-November-2018].}


\bibitem[\protect\citeauthoryear{Kulkarni, Devi, Sirighee, Hazra, and
  Rao}{Kulkarni et~al\mbox{.}}{2018}]%
        {kulkarni2018predictive}
\bibfield{author}{\bibinfo{person}{Kedar Kulkarni},
  \bibinfo{person}{Umamaheswari Devi}, \bibinfo{person}{Amith Sirighee},
  \bibinfo{person}{Jagabondhu Hazra}, {and} \bibinfo{person}{Praveen Rao}.}
  \bibinfo{year}{2018}\natexlab{}.
\newblock \showarticletitle{Predictive Maintenance for Supermarket
  Refrigeration Systems Using Only Case Temperature Data}. In
  \bibinfo{booktitle}{\emph{2018 Annual American Control Conference (ACC)}}.
  \bibinfo{publisher}{IEEE}, \bibinfo{address}{Piscataway, NJ, USA},
  \bibinfo{pages}{4640--4645}.
\newblock


\bibitem[\protect\citeauthoryear{Li}{Li}{2004}]%
        {li2004decoupling}
\bibfield{author}{\bibinfo{person}{Haorong Li}.}
  \bibinfo{year}{2004}\natexlab{}.
\newblock \emph{\bibinfo{title}{A decoupling-based unified fault detection and
  diagnosis approach for packaged air conditioners}}.
\newblock \bibinfo{thesistype}{Ph.D. Dissertation}. \bibinfo{school}{Purdue
  University}.
\newblock


\bibitem[\protect\citeauthoryear{Li, Liu, Lau, and Zhang}{Li
  et~al\mbox{.}}{2014}]%
        {li2014experimental}
\bibfield{author}{\bibinfo{person}{Yunhua Li}, \bibinfo{person}{Mingsheng Liu},
  \bibinfo{person}{Josephine Lau}, {and} \bibinfo{person}{Bei Zhang}.}
  \bibinfo{year}{2014}\natexlab{}.
\newblock \showarticletitle{Experimental study on electrical signatures of
  common faults for packaged DX rooftop units}.
\newblock \bibinfo{journal}{\emph{Energy and Buildings}}  \bibinfo{volume}{77}
  (\bibinfo{year}{2014}), \bibinfo{pages}{401--415}.
\newblock


\bibitem[\protect\citeauthoryear{Lu, Sookoor, Srinivasan, Gao, Holben,
  Stankovic, Field, and Whitehouse}{Lu et~al\mbox{.}}{2010}]%
        {lu2010smart}
\bibfield{author}{\bibinfo{person}{Jiakang Lu}, \bibinfo{person}{Tamim
  Sookoor}, \bibinfo{person}{Vijay Srinivasan}, \bibinfo{person}{Ge Gao},
  \bibinfo{person}{Brian Holben}, \bibinfo{person}{John Stankovic},
  \bibinfo{person}{Eric Field}, {and} \bibinfo{person}{Kamin Whitehouse}.}
  \bibinfo{year}{2010}\natexlab{}.
\newblock \showarticletitle{The Smart Thermostat: Using Occupancy Sensors to
  Save Energy in Homes}. In \bibinfo{booktitle}{\emph{Proceedings of the 8th
  ACM Conference on Embedded Networked Sensor Systems}}.
  \bibinfo{publisher}{ACM}, \bibinfo{address}{New York, NY, USA},
  \bibinfo{pages}{211--224}.
\newblock


\bibitem[\protect\citeauthoryear{Martell and Krcma}{Martell and Krcma}{1994}]%
        {martell1994refrigerant}
\bibfield{author}{\bibinfo{person}{Dennis Martell} {and} \bibinfo{person}{Jan
  Krcma}.} \bibinfo{year}{1994}\natexlab{}.
\newblock \bibinfo{title}{Refrigerant gas leak detector}.
\newblock
\newblock
\newblock
\shownote{US Patent 5,351,037.}


\bibitem[\protect\citeauthoryear{Mavromatidis, Acha, and Shah}{Mavromatidis
  et~al\mbox{.}}{2013}]%
        {mavromatidis2013diagnostic}
\bibfield{author}{\bibinfo{person}{Georgios Mavromatidis},
  \bibinfo{person}{Salvador Acha}, {and} \bibinfo{person}{Nilay Shah}.}
  \bibinfo{year}{2013}\natexlab{}.
\newblock \showarticletitle{Diagnostic tools of energy performance for
  supermarkets using Artificial Neural Network algorithms}.
\newblock \bibinfo{journal}{\emph{Energy and Buildings}}  \bibinfo{volume}{62}
  (\bibinfo{year}{2013}), \bibinfo{pages}{304--314}.
\newblock


\bibitem[\protect\citeauthoryear{Meng, Dunham, Marchetti, and Huang}{Meng
  et~al\mbox{.}}{2006}]%
        {meng2006rare}
\bibfield{author}{\bibinfo{person}{Yu Meng}, \bibinfo{person}{Margaret~H
  Dunham}, \bibinfo{person}{F~Marco Marchetti}, {and} \bibinfo{person}{Jie
  Huang}.} \bibinfo{year}{2006}\natexlab{}.
\newblock \showarticletitle{Rare event detection in a spatiotemporal
  environment}. In \bibinfo{booktitle}{\emph{Granular Computing, 2006 IEEE
  International Conference on}}. \bibinfo{publisher}{IEEE},
  \bibinfo{address}{Piscataway, NJ, USA}, \bibinfo{pages}{629--634}.
\newblock


\bibitem[\protect\citeauthoryear{Morrow}{Morrow}{1994}]%
        {morrow1994refrigerant}
\bibfield{author}{\bibinfo{person}{Gordon~R Morrow}.}
  \bibinfo{year}{1994}\natexlab{}.
\newblock \bibinfo{title}{Refrigerant leak detector system}.
\newblock
\newblock
\newblock
\shownote{US Patent 5,351,500.}


\bibitem[\protect\citeauthoryear{Morrow and White}{Morrow and White}{1996}]%
        {morrow1996refrigerant}
\bibfield{author}{\bibinfo{person}{Gordon~R Morrow} {and}
  \bibinfo{person}{Coy~M White}.} \bibinfo{year}{1996}\natexlab{}.
\newblock \bibinfo{title}{Refrigerant leak detector system}.
\newblock
\newblock
\newblock
\shownote{US Patent 5,524,445.}


\bibitem[\protect\citeauthoryear{Narayanaswamy, Balaji, Gupta, and
  Agarwal}{Narayanaswamy et~al\mbox{.}}{2014}]%
        {narayanaswamy2014data}
\bibfield{author}{\bibinfo{person}{Balakrishnan Narayanaswamy},
  \bibinfo{person}{Bharathan Balaji}, \bibinfo{person}{Rajesh Gupta}, {and}
  \bibinfo{person}{Yuvraj Agarwal}.} \bibinfo{year}{2014}\natexlab{}.
\newblock \showarticletitle{Data Driven Investigation of Faults in HVAC Systems
  with Model, Cluster and Compare (MCC)}. In
  \bibinfo{booktitle}{\emph{Proceedings of the 1st ACM Conference on Embedded
  Systems for Energy-Efficient Buildings}}. \bibinfo{publisher}{ACM},
  \bibinfo{address}{New York, NY, USA}, \bibinfo{pages}{50--59}.
\newblock


\bibitem[\protect\citeauthoryear{Navarro-Esbri, Torrella, and
  Cabello}{Navarro-Esbri et~al\mbox{.}}{2006}]%
        {navarro2006vapour}
\bibfield{author}{\bibinfo{person}{J Navarro-Esbri}, \bibinfo{person}{Eb
  Torrella}, {and} \bibinfo{person}{R Cabello}.}
  \bibinfo{year}{2006}\natexlab{}.
\newblock \showarticletitle{A vapour compression chiller fault detection
  technique based on adaptative algorithms. Application to on-line refrigerant
  leakage detection}.
\newblock \bibinfo{journal}{\emph{International Journal of Refrigeration}}
  \bibinfo{volume}{29} (\bibinfo{year}{2006}), \bibinfo{pages}{716--723}.
\newblock


\bibitem[\protect\citeauthoryear{O'Neill, Pang, Shashanka, Haves, and
  Bailey}{O'Neill et~al\mbox{.}}{2014}]%
        {o2014model}
\bibfield{author}{\bibinfo{person}{Zheng O'Neill}, \bibinfo{person}{Xiufeng
  Pang}, \bibinfo{person}{Madhusudana Shashanka}, \bibinfo{person}{Philip
  Haves}, {and} \bibinfo{person}{Trevor Bailey}.}
  \bibinfo{year}{2014}\natexlab{}.
\newblock \showarticletitle{Model-based real-time whole building energy
  performance monitoring and diagnostics}.
\newblock \bibinfo{journal}{\emph{Journal of Building Performance Simulation}}
  \bibinfo{volume}{7} (\bibinfo{year}{2014}), \bibinfo{pages}{83--99}.
\newblock


\bibitem[\protect\citeauthoryear{Palani, Nasir, Prakash, Chugh, Gupta, and
  Ramamritham}{Palani et~al\mbox{.}}{2014}]%
        {palani2014putting}
\bibfield{author}{\bibinfo{person}{Kartik Palani}, \bibinfo{person}{Nabeel
  Nasir}, \bibinfo{person}{Vivek~Chil Prakash}, \bibinfo{person}{Amandeep
  Chugh}, \bibinfo{person}{Rohit Gupta}, {and} \bibinfo{person}{Krithi
  Ramamritham}.} \bibinfo{year}{2014}\natexlab{}.
\newblock \showarticletitle{Putting Smart Meters to Work: Beyond the Usual}. In
  \bibinfo{booktitle}{\emph{Proceedings of the 5th International Conference on
  Future Energy Systems}}. \bibinfo{publisher}{ACM}, \bibinfo{address}{New
  York, NY, USA}, \bibinfo{pages}{237--238}.
\newblock


\bibitem[\protect\citeauthoryear{Parekh}{Parekh}{1992}]%
        {parekh1992method}
\bibfield{author}{\bibinfo{person}{Manher Parekh}.}
  \bibinfo{year}{1992}\natexlab{}.
\newblock \bibinfo{title}{Method for detecting leakage in a refrigeration
  system}.
\newblock
\newblock
\newblock
\shownote{US Patent 5,149,453.}


\bibitem[\protect\citeauthoryear{Payne, Domanski, Heo, and Du}{Payne
  et~al\mbox{.}}{2015}]%
        {payne2015self}
\bibfield{author}{\bibinfo{person}{William~V Payne}, \bibinfo{person}{Piotr~A
  Domanski}, \bibinfo{person}{Jaehyeok Heo}, {and} \bibinfo{person}{Zhimin
  Du}.} \bibinfo{year}{2015}\natexlab{}.
\newblock \bibinfo{booktitle}{\emph{Self-Training of a Fault-Free Model for
  Residential Air Conditioner Fault Detection and Diagnostics}}.
\newblock \bibinfo{type}{{T}echnical {R}eport}. \bibinfo{institution}{NIST,
  USA}.
\newblock


\bibitem[\protect\citeauthoryear{Payne, Heo, and Domanski}{Payne
  et~al\mbox{.}}{2018}]%
        {payne2018data}
\bibfield{author}{\bibinfo{person}{W~Vance Payne}, \bibinfo{person}{Jaehyeok
  Heo}, {and} \bibinfo{person}{Piotr~A Domanski}.}
  \bibinfo{year}{2018}\natexlab{}.
\newblock \showarticletitle{A Data-Clustering Technique for Fault Detection and
  Diagnostics in Field-Assembled Air Conditioners}.
\newblock \bibinfo{journal}{\emph{International Journal of Air-Conditioning and
  Refrigeration}}  \bibinfo{volume}{26} (\bibinfo{year}{2018}),
  \bibinfo{pages}{1850015}.
\newblock


\bibitem[\protect\citeauthoryear{Pedregosa, Varoquaux, Gramfort, Michel,
  Thirion, Grisel, Blondel, Prettenhofer, Weiss, Dubourg,
  et~al\mbox{.}}{Pedregosa et~al\mbox{.}}{2011}]%
        {pedregosa2011scikit}
\bibfield{author}{\bibinfo{person}{Fabian Pedregosa}, \bibinfo{person}{Ga{\"e}l
  Varoquaux}, \bibinfo{person}{Alexandre Gramfort}, \bibinfo{person}{Vincent
  Michel}, \bibinfo{person}{Bertrand Thirion}, \bibinfo{person}{Olivier
  Grisel}, \bibinfo{person}{Mathieu Blondel}, \bibinfo{person}{Peter
  Prettenhofer}, \bibinfo{person}{Ron Weiss}, \bibinfo{person}{Vincent
  Dubourg}, {et~al\mbox{.}}} \bibinfo{year}{2011}\natexlab{}.
\newblock \showarticletitle{Scikit-learn: Machine learning in Python}.
\newblock \bibinfo{journal}{\emph{Journal of machine learning research}}
  \bibinfo{volume}{12} (\bibinfo{year}{2011}), \bibinfo{pages}{2825--2830}.
\newblock


\bibitem[\protect\citeauthoryear{Porter, Rozsnaki, and Ahmed}{Porter
  et~al\mbox{.}}{2008}]%
        {porter2008refrigeration}
\bibfield{author}{\bibinfo{person}{Michael~Ramey Porter},
  \bibinfo{person}{Joseph~James Rozsnaki}, {and} \bibinfo{person}{Osman
  Ahmed}.} \bibinfo{year}{2008}\natexlab{}.
\newblock \bibinfo{title}{Refrigeration system fault detection and diagnosis
  using distributed microsystems}.
\newblock
\newblock
\newblock
\shownote{US Patent App. 11/786,033.}


\bibitem[\protect\citeauthoryear{Ren, Liang, Gu, and Han}{Ren
  et~al\mbox{.}}{2008}]%
        {ren2008fault}
\bibfield{author}{\bibinfo{person}{Neng Ren}, \bibinfo{person}{Jun Liang},
  \bibinfo{person}{Bo Gu}, {and} \bibinfo{person}{Hua Han}.}
  \bibinfo{year}{2008}\natexlab{}.
\newblock \showarticletitle{Fault diagnosis strategy for incompletely described
  samples and its application to refrigeration system}.
\newblock \bibinfo{journal}{\emph{Mechanical Systems and Signal Processing}}
  \bibinfo{volume}{22} (\bibinfo{year}{2008}), \bibinfo{pages}{436--450}.
\newblock


\bibitem[\protect\citeauthoryear{Rinehart}{Rinehart}{2004}]%
        {rinehart2004refrigerant}
\bibfield{author}{\bibinfo{person}{Charlie~M Rinehart}.}
  \bibinfo{year}{2004}\natexlab{}.
\newblock \bibinfo{title}{Refrigerant leak detection system}.
\newblock
\newblock
\newblock
\shownote{US Patent 6,772,598.}


\bibitem[\protect\citeauthoryear{Rossi and Braun}{Rossi and Braun}{1997}]%
        {rossi1997statistical}
\bibfield{author}{\bibinfo{person}{Todd~M Rossi} {and} \bibinfo{person}{James~E
  Braun}.} \bibinfo{year}{1997}\natexlab{}.
\newblock \showarticletitle{A statistical, rule-based fault detection and
  diagnostic method for vapor compression air conditioners}.
\newblock \bibinfo{journal}{\emph{Hvac\&R Research}} \bibinfo{volume}{3},
  \bibinfo{number}{1} (\bibinfo{year}{1997}), \bibinfo{pages}{19--37}.
\newblock


\bibitem[\protect\citeauthoryear{{Sense}}{{Sense}}{2018}]%
        {Sense}
\bibfield{author}{\bibinfo{person}{{Sense}}.} \bibinfo{year}{2018}\natexlab{}.
\newblock \bibinfo{title}{The Sense Home Energy Monitor}.
\newblock
\newblock
\urldef\tempurl%
\url{https://sense.com}
\showURL{%
\tempurl}
\newblock
\shownote{[Online; accessed 12-November-2018].}


\bibitem[\protect\citeauthoryear{Srinivasan, Vasan, Sarangan, and
  Sivasubramaniam}{Srinivasan et~al\mbox{.}}{2015}]%
        {srinivasan2015bugs}
\bibfield{author}{\bibinfo{person}{Shravan Srinivasan},
  \bibinfo{person}{Arunchandar Vasan}, \bibinfo{person}{Venkatesh Sarangan},
  {and} \bibinfo{person}{Anand Sivasubramaniam}.}
  \bibinfo{year}{2015}\natexlab{}.
\newblock \showarticletitle{Bugs in the Freezer: Detecting Faults in
  Supermarket Refrigeration Systems Using Energy Signals}. In
  \bibinfo{booktitle}{\emph{Proceedings of the 2015 ACM Sixth International
  Conference on Future Energy Systems}}. \bibinfo{publisher}{ACM},
  \bibinfo{address}{New York, NY, USA}, \bibinfo{pages}{101--110}.
\newblock


\bibitem[\protect\citeauthoryear{Suzuki, Tsunooka, Tahara, and Nobuta}{Suzuki
  et~al\mbox{.}}{2004}]%
        {suzuki2004vapor}
\bibfield{author}{\bibinfo{person}{Takahisa Suzuki}, \bibinfo{person}{Tatsuo
  Tsunooka}, \bibinfo{person}{Toshihiro Tahara}, {and} \bibinfo{person}{Tetsuji
  Nobuta}.} \bibinfo{year}{2004}\natexlab{}.
\newblock \bibinfo{title}{Vapor compression type refrigeration apparatus
  including leak detection and method for detecting refrigerant leaks}.
\newblock
\newblock
\newblock
\shownote{US Patent 6,681,582.}


\bibitem[\protect\citeauthoryear{Tassou and Grace}{Tassou and Grace}{2005}]%
        {tassou2005fault}
\bibfield{author}{\bibinfo{person}{SA Tassou} {and} \bibinfo{person}{IN
  Grace}.} \bibinfo{year}{2005}\natexlab{}.
\newblock \showarticletitle{Fault diagnosis and refrigerant leak detection in
  vapour compression refrigeration systems}.
\newblock \bibinfo{journal}{\emph{International Journal of Refrigeration}}
  \bibinfo{volume}{28} (\bibinfo{year}{2005}), \bibinfo{pages}{680--688}.
\newblock


\bibitem[\protect\citeauthoryear{Taylor and Corne}{Taylor and Corne}{2004}]%
        {taylor2004refrigerant}
\bibfield{author}{\bibinfo{person}{Dan~W Taylor} {and} \bibinfo{person}{David~W
  Corne}.} \bibinfo{year}{2004}\natexlab{}.
\newblock \showarticletitle{Refrigerant leak prediction in supermarkets using
  evolved neural networks}.
\newblock In \bibinfo{booktitle}{\emph{Recent Advances In Simulated Evolution
  And Learning}}. \bibinfo{publisher}{World Scientific},
  \bibinfo{address}{Hackensack, NJ, USA}.
\newblock


\bibitem[\protect\citeauthoryear{Thybo and Izadi-Zamanabadi}{Thybo and
  Izadi-Zamanabadi}{2004}]%
        {thybo2004development}
\bibfield{author}{\bibinfo{person}{Claus Thybo} {and} \bibinfo{person}{Roozbeh
  Izadi-Zamanabadi}.} \bibinfo{year}{2004}\natexlab{}.
\newblock \showarticletitle{Development of fault detection and diagnosis
  schemes for industrial refrigeration systems-Lessons learned}. In
  \bibinfo{booktitle}{\emph{Control Applications, 2004. Proceedings of the 2004
  IEEE International Conference on}}, Vol.~\bibinfo{volume}{2}.
  \bibinfo{publisher}{IEEE}, \bibinfo{address}{Piscataway, NJ, USA},
  \bibinfo{pages}{1248--1253}.
\newblock


\bibitem[\protect\citeauthoryear{{TWC Product and Technology}}{{TWC Product and
  Technology}}{2014}]%
        {wunderground}
\bibfield{author}{\bibinfo{person}{{TWC Product and Technology}}.}
  \bibinfo{year}{2014}\natexlab{}.
\newblock \bibinfo{title}{Wunderground}.
\newblock
\newblock
\urldef\tempurl%
\url{http://www.wunderground.com/}
\showURL{%
\tempurl}
\newblock
\shownote{[Online; accessed 12-November-2018].}


\bibitem[\protect\citeauthoryear{{Wikipedia contributors}}{{Wikipedia
  contributors}}{2017}]%
        {wikipedia2018cusum}
\bibfield{author}{\bibinfo{person}{{Wikipedia contributors}}.}
  \bibinfo{year}{2017}\natexlab{}.
\newblock \bibinfo{title}{CUSUM --- {Wikipedia}{,} The Free Encyclopedia}.
\newblock
\newblock
\urldef\tempurl%
\url{https://en.wikipedia.org/w/index.php?title=CUSUM&oldid=786593468}
\showURL{%
\tempurl}
\newblock
\shownote{[Online; accessed 12-November-2018].}


\bibitem[\protect\citeauthoryear{Wirz}{Wirz}{2017}]%
        {wirz2017commercial}
\bibfield{author}{\bibinfo{person}{Dick Wirz}.}
  \bibinfo{year}{2017}\natexlab{}.
\newblock \bibinfo{booktitle}{\emph{Commercial refrigeration for air
  conditioning technicians}}.
\newblock \bibinfo{publisher}{Cengage Learning}, \bibinfo{address}{Boston,
  USA}.
\newblock


\bibitem[\protect\citeauthoryear{Yang, Rasmussen, Kieu, and
  Izadi-Zamanabadi}{Yang et~al\mbox{.}}{2011a}]%
        {yang2011afault}
\bibfield{author}{\bibinfo{person}{Zhenyu Yang}, \bibinfo{person}{Karsten~B
  Rasmussen}, \bibinfo{person}{Anh~T Kieu}, {and} \bibinfo{person}{Roozbeh
  Izadi-Zamanabadi}.} \bibinfo{year}{2011}\natexlab{a}.
\newblock \showarticletitle{Fault Detection and Isolation for a Supermarket
  Refrigeration System--Part One: Kalman-Filter-Based Methods}.
\newblock \bibinfo{journal}{\emph{IFAC Proceedings Volumes}}
  \bibinfo{volume}{44} (\bibinfo{year}{2011}), \bibinfo{pages}{13233--13238}.
\newblock


\bibitem[\protect\citeauthoryear{Yang, Rasmussen, Kieu, and
  Izadi-Zamanabadi}{Yang et~al\mbox{.}}{2011b}]%
        {yang2011bfault}
\bibfield{author}{\bibinfo{person}{Zhenyu Yang}, \bibinfo{person}{Karsten~B
  Rasmussen}, \bibinfo{person}{Anh~T Kieu}, {and} \bibinfo{person}{Roozbeh
  Izadi-Zamanabadi}.} \bibinfo{year}{2011}\natexlab{b}.
\newblock \showarticletitle{Fault Detection and Isolation for a Supermarket
  Refrigeration System--Part Two: Unknown-Input-Observer Method and Its
  Extension}.
\newblock \bibinfo{journal}{\emph{IFAC Proceedings Volumes}}
  \bibinfo{volume}{44} (\bibinfo{year}{2011}), \bibinfo{pages}{4238--4243}.
\newblock


\bibitem[\protect\citeauthoryear{Yoo, Hong, and Kim}{Yoo et~al\mbox{.}}{2017}]%
        {yoo2017refrigerant}
\bibfield{author}{\bibinfo{person}{Jin~Woo Yoo}, \bibinfo{person}{Sung~Bin
  Hong}, {and} \bibinfo{person}{Min~Soo Kim}.} \bibinfo{year}{2017}\natexlab{}.
\newblock \showarticletitle{Refrigerant leakage detection in an EEV installed
  residential air conditioner with limited sensor installations}.
\newblock \bibinfo{journal}{\emph{International Journal of Refrigeration}}
  \bibinfo{volume}{78} (\bibinfo{year}{2017}), \bibinfo{pages}{157--165}.
\newblock


\bibitem[\protect\citeauthoryear{Yu, Woradechjumroen, and Yu}{Yu
  et~al\mbox{.}}{2014}]%
        {yu2014review}
\bibfield{author}{\bibinfo{person}{Yuebin Yu}, \bibinfo{person}{Denchai
  Woradechjumroen}, {and} \bibinfo{person}{Daihong Yu}.}
  \bibinfo{year}{2014}\natexlab{}.
\newblock \showarticletitle{A review of fault detection and diagnosis
  methodologies on air-handling units}.
\newblock \bibinfo{journal}{\emph{Energy and Buildings}}  \bibinfo{volume}{82}
  (\bibinfo{year}{2014}), \bibinfo{pages}{550--562}.
\newblock


\end{thebibliography}

\end{document}